\documentclass[aps,prd,eqsecnum,11pt,showpacs,preprintnumbers]{revtex4}
 
\def\nbox#1#2{\vcenter{\hrule \hbox{\vrule height#2in
\kern#1in \vrule} \hrule}}
\def\sq{\,\raise.5pt\hbox{$\nbox{.09}{.09}$}\,} 
\def\sqb{\,\raise.5pt\hbox{$\overline{\nbox{.09}{.09}}$}\,}

\usepackage{amssymb,amsmath,subfig}
\usepackage{graphicx}

\begin{document}

\preprint{\rm LA-UR-06-2112}

\title{Macroscopic Effects of the Quantum Trace Anomaly}
\author{Emil Mottola\footnote{E-mail: emil@lanl.gov}}
\affiliation{Theoretical Division, T-8 \\
Los Alamos National Laboratory \\
M.S. B285 \\
Los Alamos, NM 87545\\}
\author{Ruslan Vaulin\footnote{E-mail: vaulin@physics.fau.edu}}
\affiliation{Dept.of Physics \\
Florida Atlantic University\\
777 Glades Road\\
Boca Raton, FL 33431}

\begin{abstract}
\vskip .5cm
The low energy effective action of gravity in any even dimension generally
acquires non-local terms associated with the trace anomaly, generated by
the quantum fluctuations of massless fields. The local auxiliary field 
description of this effective action in four dimensions requires two 
additional scalar fields, not contained in classical general relativity, 
which remain relevant at macroscopic distance scales. The auxiliary 
scalar fields depend upon boundary conditions for their complete 
specification, and therefore carry global information about the geometry 
and macroscopic quantum state of the gravitational field. The scalar 
potentials also provide coordinate invariant order parameters describing 
the conformal behavior and divergences of the stress tensor on event horizons. 
We compute the stress tensor due to the anomaly in terms of its auxiliary 
scalar potentials in a number of concrete examples, including the Rindler 
wedge, the Schwarzschild geometry, and de Sitter spacetime. In all of 
these cases, a small number of classical order parameters completely 
determine the divergent behaviors allowed on the horizon, and yield 
qualitatively correct global approximations to the renormalized 
expectation value of the quantum stress tensor.\\
\end{abstract} 

        
\maketitle
 
\vfill\eject

\section{Introduction}

Gravitation is a macroscopic phenomenon, governing the behavior of matter
at scales from a few cm. up to the largest scales of cosmology, $10^{28}$ cm.
On the other hand quantum phenomena are associated generally with microscopic 
distances, on the order of atomic scales, $10^{-8}$ cm. or smaller. The physical 
and mathematical frameworks used to describe physics at these disparate scales are 
quite different as well, and at first sight appear even to be irreconcilable. 
In general relativity gravitation is founded on the Equivalence Principle, which 
finds its mathematical expression in differential equations which transform 
covariantly under a local change of frame at every spacetime point. In classical
physics this idealization of spacetime as a local differentiable structure is quite 
natural. Yet because of quantum fluctuations it is very likely that the classical
description of spacetime as a smooth pseudo-Riemannian manifold fails at very 
short distance scales, of the order of the Planck length, $L_{Pl} = \sqrt{\hbar G/c^3}
\sim 10^{-33}$ cm.

Perhaps less widely appreciated is that quantum theory also raises questions 
about the purely local, classical description of the gravitational interactions of
matter, even at macroscopic distances. This is because of the peculiarly quantum 
phenomena of phase coherence and entanglement, which may be present on any scale,
given the right conditions. Indeed macroscopic quantum states are encountered 
in virtually all branches of physics on a very wide variety of scales. Some of the 
better known examples occur in non-relativistic many-body and condensed matter 
systems, such as Bose-Einstein condensation, superfluidity, and superconductivity. 
These quantum coherence effects due to the wave-like properties of matter have their 
close analogs in relativistic field theories as well, in spontaneous symmetry 
breaking by the Higgs mechanism in electroweak theory, and chiral quark pair 
condensation in low energy QCD. In each of these cases the ground state or 
``vacuum" is a macroscopic quantum state, described at very long wavelengths by a 
non-vanishing quasi-classical order parameter, in a low energy effective 
field theory.

The strictly classical theory of general relativity would seem to preclude any 
incorporation of macroscopic coherence and entanglement effects of quantum 
matter, on any scale. Taking account of the quantum wave-like properties of 
matter, and its propensity to form phase correlated states over macroscopic distance 
scales at low enough temperatures and/or high enough densities requires at the least 
a semi-classical treatment of the effective stress-energy tensor source for Einstein's
equations. The corresponding one-loop effective action for gravity can be determined 
by the methods of effective field theory (EFT). Our principal purpose in this paper 
is to demonstrate that non-local macroscopic coherence effects are contained in the 
low energy EFT of gravity, provided that Einstein's classical theory is {\it supplemented}
by the contributions of the one-loop quantum trace anomaly of massless fields.

The low energy EFT of gravity is determined by the same general principles as 
in other contexts \cite{DGH}, namely by an expansion in powers of derivatives of 
local terms consistent with symmetry. Short distance effects are parameterized 
by the coefficients of local operators in the effective action, with higher order 
terms suppressed by inverse powers of an ultraviolet cutoff scale $M$. The effective 
theory need not be renormalizable, as indeed Einstein's theory is not, but is 
expected nonetheless to be quite insensitive to the details of the underlying 
microscopic degrees of freedom, because of decoupling \cite{DGH}. It is the 
decoupling of short distance degrees of freedom from the macroscopic physics that 
makes EFT techniques so widely applicable, and which we assume applies also to gravity.

As a covariant metric theory with a symmetry dictated by the Equivalence Principle, 
namely invariance under general coordinate transformations, general relativity may 
be regarded as just such a local EFT, truncated at second order in derivatives of 
the metric field $g_{ab}(x)$ \cite{Dono}. When quantum matter is considered, 
the stress tensor becomes an operator whose expectation value acts as a source for the 
Einstein equations in the semi-classical limit. Fourth order terms in derivatives of 
the metric are necessary to absorb divergences in the expectation value of the stress 
tensor in curved spacetime, but the effects of such higher derivative {\it local} 
terms in the gravitational effective action are suppressed at distance scales
$L \gg L_{Pl}$ in the low energy EFT limit. Hence surveying only local curvature 
terms, it is often tacitly assumed that Einstein's theory contains all the low
energy macroscopic degrees of freedom of gravity, and that general relativity 
cannot be modified at macroscopic distance scales, much greater than $L_{Pl}$, 
without violating general coordinate invariance and/or EFT principles.
However, this presumption should be re-examined in the presence of 
quantum anomalies.

When a classical symmetry is broken by a quantum anomaly, the naive decoupling of 
short and long distance physics assumed by an expansion in local operators with 
ascending inverse powers of $M$ fails. In this situation even the low energy 
symmetries of the effective theory are changed by the presence of the anomaly, 
and some remnant of the ultraviolet physics survives in the low energy description. 
An anomaly can have significant effects in the low energy EFT because it is not 
suppressed by any large energy cutoff scale, surviving even in the limit 
$M \rightarrow \infty$. Any explicit breaking of the symmetry in the classical 
Lagrangian serves only to mask the effects of the anomaly, but in the right 
circumstances the effects of the non-local anomaly may still dominate the local 
terms. A well known example is the chiral anomaly in QCD with massless quarks, 
whose effects are unsuppressed by any inverse power of the EFT ultraviolet cutoff 
scale $M \sim \Lambda_{QCD}$. Although the quark masses are non-zero, and chiral 
symmetry is only approximate in nature, the chiral anomaly gives the dominant 
contribution to the low energy decay amplitude of $\pi^0\rightarrow 2\gamma$ in 
the standard model \cite{Adler,BGMF}, a contribution that is missed entirely by a 
local EFT expansion in pion fields. Instead the existence of the chiral 
anomaly requires the explicit addition to the local effective action of a {\it non-local} 
term in four physical dimensions to account for its effects \cite{WZ,DGH}.

Although when an anomaly is present, naive decoupling between the short and long 
distance degrees of freedom fails, it does so in a well-defined way, with
a coefficient that depends only on the quantum numbers of the underlying microscopic 
theory. In fact, since the chiral anomaly depends on the color charge assignments of 
the short distance quark degrees of freedom, the measured low energy decay width of
$\pi^0\rightarrow 2\gamma$ affords a clean, non-trivial test of the underlying 
microscopic quantum theory of QCD with three colors of fractionally charged 
quarks \cite{BGMF,DGH,Wein}. The bridge between short and long distance physics 
which anomalies provide is the basis for the anomaly matching conditions \cite{tHooft}.  
 
In curved space an anomaly closely related to the chiral anomaly also 
appears in massless quantum field theory \cite{anom,BirDav}. This conformal or trace 
anomaly provides us with additional infrared relevant terms that do not decouple
in the limit $M_{Pl}^{-1} = L_{Pl} \rightarrow 0$, and which should be added to 
the Einstein-Hilbert action of classical relativity, to complete the EFT 
of low energy gravity. 

That the trace anomaly terms are necessary for the low energy completion
of the EFT of gravity may be seen from the classification of possible
terms in the gravitational effective action according to their behavior under 
global Weyl rescalings \cite{MazMot}. The terms in the classical Einstein-Hilbert 
action scale with positive powers ($\sim L^4$ and $\sim L^2$) under rescaling of 
distance, and are clearly relevant operators of the low energy description. The 
non-local anomalous terms scale logarithmically ($\sim\log L$) with distance under 
Weyl rescalings, and therefore should be considered {\it marginally relevant}. Unlike 
local higher derivative terms in the effective action, which are either neutral 
or scale with negative powers of $L$ under global Weyl rescalings of the metric, 
the anomalous terms cannot be discarded in the low energy, large distance limit. 
The addition of the anomaly term(s) to the low energy effective action of gravity 
amounts to a non-trivial infrared modification of general relativity, fully 
consistent with both quantum theory and the Equivalence Principle \cite{MazMot}. 

The anomalous terms in the effective action lead to corresponding additional
terms in the stress tensor and equations of motion of the low energy EFT
of gravity. These terms are most conveniently expressed in terms of
two auxiliary scalar fields which permit the non-local effective action of
the trace anomaly and its variations to be cast into local form. The
additional scalar degrees of freedom in these terms take account of 
macroscopic effects of quantum matter in gravitational fields, which
are not contained in the purely classical, local metric description of 
Einstein's theory.

In this paper we give explicit formulae for the stress tensor due to the trace 
anomaly of massless fields to be added to the classical Einstein equations in the 
low energy EFT limit, and expose its macroscopic effects in a number of familiar 
geometries. The most interesting effects are associated with geometries 
that have horizons, boundaries, or non-trivial global topologies. For example,
the two new auxiliary fields in the effective action generically diverge as the 
apparent horizon of Rindler, Schwarzschild, and de Sitter spacetimes is approached. 
This behavior of the scalar auxiliary fields gives a coordinate invariant 
semi-classical order parameter description of possible divergences of stress 
tensors of quantum field theories on spacetime horizons, enabling an understanding
of such divergences in terms of the global properties of the geometry.
The Casimir energy density between two conducting plates is an example 
of a bounded space without horizons or curvature which also admits a simple
description in terms of the scalar auxiliary fields. The various examples 
taken together suggest that the degrees of freedom contained in the scalar 
potentials induced by the trace anomaly should be regarded as semi-classical 
macroscopic order parameter fields (condensates), whose non-vanishing values 
are connected with non-trivial boundary conditions, horizons, or the topology of 
spacetime.

The paper is organized as follows. In the next section we review the auxiliary 
field form of the effective action of the trace anomaly in two dimensions, 
recovering results for the stress tensor of conformal fields in the two dimensional 
Rindler wedge, Schwarzschild and de Sitter metrics. In Section III we discuss the 
general form of the effective action in four dimensions, and give the additional
terms in the stress tensor in terms of the auxiliary fields, as well as the conserved 
Noether current associated with it. In Section IV we consider the effects of
these additional terms and the new local scalar degrees of freedom of low
energy gravity they represent in flat, Rindler and the conformally flat de Sitter
spacetime. In Section V we discuss the approximate conformal symmetry which
pertains in the vicinity of any static, spherically symmetric Killing horizon
and give the explicit form of the auxiliary fields in the Schwarzschild case. 
We characterize the possible singularities of the stress tensor on the Schwarzschild 
horizon, and find approximate stress tensors for the Hartle-Hawking, Boulware, and 
Unruh states, comparing our results with numerical results for 
$\langle T_a^{\ b}\rangle$ of massless fields of various spins. We conclude 
in Section VI with a summary of our results and their implications for quantum 
effects in gravity. The Appendix contains the explicit formulae for the
components of the stress-energy in static spherically symmetric spacetimes,
upon which the Schwarzschild and de Sitter results are based.

\section{Effective Action and Stress Tensor in Two Dimensions}

The effective action for the trace anomaly in any even dimension
is non-local when expressed in terms only of the metric $g_{ab}$.
It can be made local by the introduction of scalar auxiliary field(s). 
To illustrate this procedure and introduce the general framework which
we will use extensively in four dimensions, we consider first the 
somewhat simpler case of two physical dimensions. In $d=2$ the
trace anomaly takes the simple form \cite{BirDav},
\begin{equation}
\langle T_a^{\ a} \rangle = \frac{N}{24\pi}\, R\,,\qquad (d=2)
\label{trtwo}
\end{equation}
where $N = N_S + N_F$ is the total number of massless fields, either
scalar ($N_S$) or fermionic ($N_F$). The fact that the anomalous
trace is independent of the quantum state of the matter field(s),
and dependent only on the geometry through the local 
Ricci scalar $R$ suggests that it should be regarded as a geometric 
effect. However, no local coordinate invariant action exists whose 
metric variation leads to (\ref{trtwo}). 

A non-local action corresponding to (\ref{trtwo}) can be found by introducing 
the conformal parameterization
of the metric,
\begin{equation}
g_{ab} = e^{2\sigma} \bar g_{ab}\,,
\label{metconf}
\end{equation}
and noticing that the scalar curvature densities of the two metrics $g_{ab}$ and
$\bar g_{ab}$ are related by
\begin{equation}
R \,\sqrt{-g} = \bar R \,\sqrt{-\bar g} - 2 \,\sqrt{-\bar g} \sqb \sigma\,,
\qquad (d=2)
\label{RRbar}
\end{equation}
a linear relation in $\sigma$ in two (and only two) dimensions. Multiplying 
(\ref{trtwo}) by $\sqrt{-g}$, using (\ref{RRbar}) and noting that 
$\sqrt{-g}\langle T_a^{\ a} \rangle$ defines the conformal variation,
$\delta \Gamma^{(2)}/\delta \sigma$ of an effective action $\Gamma^{(2)}$, we conclude 
that the $\sigma$ dependence of $\Gamma^{(2)}$ can be at most quadratic in $\sigma$. 
Hence the Wess-Zumino effective action in two dimensions, $\Gamma_{WZ}^{(2)}$ is
\begin{equation}
\Gamma_{WZ}^{(2)} [\bar g ; \sigma ] = \frac{N}{24\pi}  \int\,d^2x\,\sqrt{-\bar g}
\left( - \sigma \sqb \sigma + \bar R\,\sigma\right)\,.
\label{WZact}
\end{equation}
This action functional of the base metric $\bar g_{ab}$ and the Weyl
shift parameter $\sigma$ may be regarded as a one-form representative of 
the cohomology of the local Weyl group in two dimensions \cite{MazMot}.
This means that $\Gamma_{WZ}^{(2)}[\bar g;\sigma]$ is closed under 
the co-boundary ({\it i.e.} anti-symmetrized) composition of Weyl shifts,
\begin{equation}
\Delta_{\sigma_2} \circ \Gamma_{WZ}^{(2)}[\bar g;\sigma_1] =
\Gamma_{WZ}^{(2)}[\bar g e^{2\sigma_1};\sigma_2] - 
\Gamma_{WZ}^{(2)}[\bar g e^{2\sigma_2};\sigma_1]
+ \Gamma_{WZ}^{(2)}[\bar g;\sigma_1] - \Gamma_{WZ}^{(2)}[\bar g;\sigma_2] = 0\,.
\label{WZcob}
\end{equation}
but is non-exact, in the sense that $\Gamma_{WZ}^{(2)}$ cannot itself be written as the 
co-boundary shift of a local action functional. The relation (\ref{WZcob}) following 
directly from the hermiticity of $\sq$ is exactly the Wess-Zumino consistency condition 
for $\Gamma_{WZ}^{(2)}[\bar g;\sigma]$ \cite{WZ,AMMc}.

Although $\Gamma_{WZ}^{(2)}$ cannot be written as a coboundary shift $\Delta_{\sigma}$ 
of a local single-valued covariant scalar functional of the metric $\bar g_{ab}$, it 
is straightforward to find a {\it non-local} scalar functional $S_{anom}[g]$ such that
\begin{equation}
\Gamma_{WZ}^{(2)} [\bar g ; \sigma ] = \Delta_\sigma \circ S_{anom}^{(2)}[\bar g] \equiv
S_{anom}^{(2)}[g] - S_{anom}^{(2)}[\bar g]\,.
\label{cohom}
\end{equation}
Indeed by solving (\ref{RRbar}) formally for $\sigma$, and using the 
fact that $\sqrt{-g} \sq = \sqrt{-\bar g} \sqb$ is conformally invariant in
two (and only two) dimensions, we find that $\Gamma_{WZ}^{(2)}$ can be written as 
a Weyl shift (\ref{cohom}) with
\begin{equation}
S_{anom}^{(2)}[g] = \frac{Q^2}{16\pi} \int\,d^2x\,\sqrt{-g}
\int\,d^2x'\,\sqrt{-g'}\, R(x)\,{\sq}^{-1}(x,x')\,R(x')\,,
\label{acttwo}
\end{equation}
and ${\sq}^{-1}(x,x')$ denoting the Green's function inverse of the scalar 
differential operator $\sq$. The parameter $Q^2$ is $-N/6$ if only matter fields 
in a fixed spacetime metric are considered. It becomes $(25 - N)/6$ if account is 
taken of the contributions of the metric fluctuations themselves in addition to 
those of the $N$ matter fields, thus effectively replacing $N$ by $N-25$ \cite{dress}. 
In the general case, the coefficient $Q^2$ is arbitrary, and can be treated as simply 
an additional free parameter of the low energy effective action, to be fixed
by experiment.

The anomalous effective action (\ref{acttwo}) is a scalar under coordinate 
transformations and therefore fully covariant and geometric in character. However 
since it involves the Green's function $\sq^{-1}(x,x')$, which requires boundary
conditions for its unique specification, it is quite non-local, and dependent
upon more than just the local curvature invariants of spacetime. The non-local 
and non-single valued functional of the metric, $S_{anom}^{(2)}$ may be expressed 
in a local form by the standard method of introducing auxiliary fields. In the 
case of (\ref{acttwo}) a single scalar auxiliary field, $\varphi$ satisfying
\begin{equation}
- \sq \varphi = R
\label{auxeomtwo}
\end{equation}
is sufficient. Indeed, varying
\begin{equation}
S_{anom}^{(2)}[g;\varphi]  \equiv \frac{Q^2}{16\pi} \int\,d^2x\,\sqrt{-g}\,
\left(\nabla_a \varphi\,\nabla^a \varphi - 2 R\,\varphi\right)
\label{actauxtwo}
\end{equation}
with respect to $\varphi$ gives the Eq. of motion (\ref{auxeomtwo})
for the auxiliary field, which when solved (formally) by $\varphi =-{\sq}^{-1}R$
and substituted back into $S_{anom}^{(2)}[g;\varphi]$ returns the non-local
form of the anomalous action (\ref{acttwo}), up to a surface term. 
The non-local information in addition to the local geometry which was previously 
contained in the specification of the Green's function ${\sq}^{-1}(x,x')$ now resides
in the auxiliary local field $\varphi (x)$, and the freedom to add to it
homogeneous solutions of (\ref{auxeomtwo}).

The variation of (\ref{actauxtwo}) with respect to the metric yields 
a stress-energy tensor,
\begin{eqnarray}
&&T_{ab}^{(2)}[g; \varphi] \equiv -\frac{2}{\sqrt{-g}} 
{\delta S_{anom}^{(2)} [g; \varphi] \over \delta g^{ab}} \nonumber\\
&& \quad = {Q^2\over 4\pi}\left[-\nabla_a\nabla_b \varphi
+ g_{ab}\, \sq\varphi - {1\over 2}(\nabla_a\varphi)(\nabla_b\varphi)
+ {1\over 4} g_{ab}\, (\nabla_c\varphi)(\nabla^c\varphi)\right]\,,
\label{anomTtwo}
\end{eqnarray}
which is covariantly conserved, by use of (\ref{auxeomtwo}) and the 
vanishing of the Einstein tensor, $G_{ab} = R_{ab} - Rg_{ab}/2 = 0$ 
in two (and only two) dimensions. The {\it classical} trace of the 
stress tensor,
\begin{equation}
g^{ab}T_{ab}^{(2)}[g; \varphi] = {Q^2\over 4\pi} \sq \varphi
=- {Q^2\over 4\pi}\,R
\label{trTtwo}
\end{equation}
reproduces the {\it quantum} trace anomaly in a general classical background
(with $Q^2$ proportional to $\hbar$). Hence (\ref{actauxtwo}) is exactly the 
local auxiliary field form of the effective action which should be added to the 
action for two dimensional gravity to take the trace anomaly of massless quantum 
fields into account.

The full effective action of two dimensional gravity is
\begin{equation}
S_{eff}^{(2)}[g, \varphi] = S_{anom}^{(2)}[g, \varphi] + S_{cl}^{(2)}[g]\,,
\label{efftwo}
\end{equation}
where $S_{cl}^{(2)}$ is the local classical Einstein-Hilbert action,
\begin{equation}
S_{cl}^{(2)}[g] = \gamma \int\, d^2x\, \sqrt{-g}\, R + \lambda \int\, d^2x\, \sqrt{-g}\,,
\label{cltwo}
\end{equation}
which are the only terms we would have written down according
to the usual EFT approach, expanding in strictly local terms up to and including 
dimension two which transform as scalars under general coordinate transformations. 
Under global rescalings of the metric (\ref{metconf}) with $\sigma = \sigma_0$, a 
spacetime constant, the volume term scales like $e^{2\sigma_0}$ and its coefficient 
$\lambda$ has positive mass dimension two. Hence it is clearly a relevant term in the 
effective action at large distances. The integral of the Ricci scalar is 
independent of $\sigma_0$, with a coefficient $\gamma$ that is dimensionless,
while the integrals of all higher local curvature invariants scale with 
negative powers of $e^{\sigma_0}$, multiplied by coefficients having negative
mass dimensions. These terms are neglected in (\ref{cltwo}) since they
are strictly irrelevant in the low energy EFT limit.

The anomalous term (\ref{acttwo}), or (\ref{actauxtwo}) scales linearly with 
$\sigma_0$, {\it i.e.} logarithmically with distance, and is therefore
a {\it marginally relevant} operator in the infrared. Even if it is not included
at the classical level it will be generated by the one-loop effects of massless 
fields, which do not decouple at any scale. Hence it should be retained 
in the full low energy effective action (\ref{efftwo}) of two dimensional gravity.
Moreover, since the integral of $R$ is a topological invariant in two dimensions, 
the classical action (\ref{cltwo}) contains no propagating degrees of freedom 
whatsoever, and it is $S_{anom}$ which contains the only kinetic terms of the 
low energy EFT. In the local auxiliary field form (\ref{actauxtwo}), it is clear 
that $S_{anom}$ describes an additional scalar degree of freedom $\varphi$, not 
contained in the classical action $S_{cl}^{(2)}$. This is reflected 
also in the shift of the central charge from $N-26$, which would be expected 
from the contribution of conformal matter plus ghosts by one unit 
to $N-25$. 

Extensive study of the stress tensor and its correlators arising from this effective 
action established that the two dimensional trace anomaly gives rise to a modification 
or gravitational ``dressing" of critical exponents in conformal field theories at 
second order critical points \cite{dress}. Since critical exponents in a second order 
phase transition depend only upon fluctuations at the largest allowed infrared scale, 
this dressing is clearly an infrared effect, independent of any ultraviolet cutoff. 
These dressed exponents are evidence of the infrared fluctuations of the additional 
scalar degree of freedom $\varphi$ which are quite absent in the classical action 
(\ref{cltwo}). The appearance of the gravitational dressing exponents and the
anomalous effective action (\ref{acttwo}) itself have been confirmed in the large 
volume scaling limit of two dimensional simplicial lattice simulations in the 
dynamical triangulation approach \cite{DT,CatMot}.

The formal similarity between (\ref{WZact}) and (\ref{actauxtwo}) suggests
that the introduction of the local auxiliary field $\varphi$ has simply undone 
the steps leading from (\ref{WZact}) to (\ref{acttwo}), which eliminated the 
conformal factor $\sigma$. By comparing (\ref{RRbar}) 
and (\ref{auxeomtwo}), we observe that the metric 
\begin{equation}
\tilde g_{ab} = e^{-\varphi} g_{ab}
\label{confac}
\end{equation}
is a metric with zero scalar curvature, $\tilde R = 0$, conformally related 
to the physical metric $g_{ab}$. Hence $\varphi(x)$ does parameterize the
local Weyl transformations of the metric along the same fiber as $\sigma(x)$ 
in the conformal parameterization (\ref{metconf}). The difference between the
two representations is that whereas $\bar g_{ab}$ was regarded as a fixed
base metric related to the physical metric $g_{ab}$ by (\ref{metconf}),
the possibility of adding homogeneous solutions to (\ref{auxeomtwo}) shows 
that $\varphi$ and hence $\tilde g_{ab}$ defined by (\ref{confac}) are
not unique, and hence no $\tilde g_{ab}$ plays any privileged role over any other.
The only physical metric is $g_{ab}$. The non-uniqueness of $\varphi$
allows for additional degrees of freedom at boundaries or coordinate singularities 
of $g_{ab}$, which need not be associated with any smooth $\tilde g_{ab}$. The 
completely covariant local form of the action (\ref{actauxtwo}) makes it clear that 
these effects, associated with the boundary conditions on the spacetime scalar 
$\varphi$ satisfying (\ref{auxeomtwo}) are not coordinate artifacts which can be 
removed by reparameterizations of coordinates in the physical spacetime $g_{ab}$.
It is the freedom to specify boundary conditions for the classical auxiliary field 
$\varphi$, or equivalently add homogeneous solutions to the non-local Green's function 
$\sq^{-1}(x,x')$ which implies the possibility of global macroscopic effects of the 
anomaly.
  
Both the auxiliary field equation of motion (\ref{auxeomtwo}) and stress 
tensor (\ref{anomTtwo}) are left invariant by the constant shift,
\begin{equation}
\varphi \rightarrow \varphi + \varphi_0\,.
\label{shift}
\end{equation}
The underlying reason for this shift symmetry is that the integral,
\begin{equation}
\chi = \frac{1}{4\pi} \int d^2 x \sqrt{-g} R
\label{Eulertwo}
\end{equation}
is a topological invariant in two dimensions. For Euclidean signature metrics 
$\chi = \chi_E$ is the Euler number of the manifold. This implies that in two 
(and only two) dimensions $\sqrt{-g} R$ can be expressed as a total derivative, 
or equivalently,
\begin{equation}
R = \frac{1}{\sqrt{-g}} \partial_a (\sqrt{-g}\, \Omega^a) \equiv \nabla_a \Omega^a 
\,,\qquad (d=2)\,.
\label{toptwo}
\end{equation}
The topological current $\Omega^a$, analogous to the Chern-Simons current for the 
gauge theory chiral anomaly \cite{ChSim,EGH}, depends upon the choice of gauge or 
coordinate frame and is non-unique, although its line integral around a closed contour,
\begin{equation}
\oint_{\partial V} \sqrt {-g}\, \Omega^a\, n_a\, ds = 
\oint_{\partial V} \sqrt {-g}\, \Omega^a\,\epsilon_{ab}\, dx^b
= \int_V \sqrt {-g}\, R\, d^2x
\end{equation}
is a coordinate invariant scalar quantity. Here we have used the
notation, $n_a = \epsilon_{ab} \frac{d x^b}{ds}$ for the normal to the boundary
$\partial V$ of the two dimensional region $V$ and $\epsilon_{ab}$
is the standard alternating tensor in two dimensions (with $\epsilon_{12} = +1$).
The existence of $\Omega^a$ satisfying (\ref{toptwo}) implies that the
action (\ref{actauxtwo}) may be rewritten in the alternative form,
\begin{equation}
S_{anom,\, 2}[g;\varphi]  = \frac{Q^2}{16\pi} \int_V\,d^2x\,\sqrt{-g}
\left(\nabla^a \varphi + 2 \Omega^a\right) \nabla_a\varphi
- \frac{Q^2}{8\pi}\oint_{\partial V}\sqrt{-g}\,\varphi\,  \Omega^a n_a ds \,,
\label{acttwoalt}
\end{equation}
leaving (\ref{auxeomtwo}), (\ref{anomTtwo}) and (\ref{trTtwo}) unchanged. 
Since up to the last surface term the anomalous action depends on 
$\varphi$ only through its derivatives, there is a frame dependent
Noether current corresponding to the global Weyl rescaling (\ref{shift}), 
\begin{equation}
J^a = \nabla^a \varphi + \Omega^a\,,
\label{Noetwo}
\end{equation}
which is covariantly conserved, 
\begin{equation}
\nabla_a J^a \equiv \frac{1}{\sqrt{-g}} \partial_a (\sqrt{-g} J^a)
= \sq \varphi + R = 0\,,
\label{Jdiv}
\end{equation}
by virtue of (\ref{auxeomtwo}) and (\ref{toptwo}). 

Within the framework of the low energy effective action (\ref{efftwo}) all fields may be 
treated classically, with $\hbar$ contained implicitly only in the c-number coefficients 
$\gamma, \lambda$ and $Q^2$. In this (semi-)classical framework, the auxiliary field 
$\varphi$ is a classical scalar potential which contains information about the 
macroscopic quantum state and topological boundary effects. The geometric information 
$\varphi(x)$ contains cannot be reduced to local geometric invariants constructed from 
the curvature tensor, its derivatives and contractions at $x$. Instead the massless 
correlation function $\sq^{-1}(x,x')$, which grows logarithmically with the invariant 
distance between the spacetime points $x$ and $x'$, and the existence of a conserved 
global charge corresponding to the Noether current $J^a$ imply that $\varphi$ is a long 
range field depending upon the global properties of spacetime. It is this long range 
behavior of $\varphi$ that enables the effective action (\ref{efftwo}) to incorporate 
macroscopic quantum coherence effects, quite absent from the purely local classical 
geometric action $S_{cl}^{(2)}$. Since the effective action contains $S_{anom}^{(2)}$,
these non-local macroscopic quantum coherence effects are incorporated in
the low energy EFT of two dimensional gravity in a natural way.

The long range macroscopic effects and the physical meaning of the
Noether charge are best illustrated by means of a few simple examples.
Consider first flat spacetime in Rindler coordinates,
\begin{equation}
ds^2 = -dt^2 + dx^2 = - \rho^2 d\eta^2 + d\rho^2,
\label{flatwo}
\end{equation}
with $x= \rho \cosh \eta,\, t = \rho \sinh \eta $. The Killing vector 
$\partial_\eta$ generates Lorentz boosts and this vector field becomes 
singular on the light cone, $\rho^2 = x^2 - t^2 = 0$. If the state of the system 
is boost invariant then the semi-classical auxiliary field can be assumed 
to be independent of $\eta$. With $\varphi = \varphi (\rho)$, (\ref{auxeomtwo}) 
has the solution (for finite non-zero $\rho$),
\begin{equation}
\varphi = 2 q \ln \left(\frac{\rho}{\rho_0}\right)\,,
\label{philog}
\end{equation}
with $q$ and $\rho_0$ arbitrary constants. 

If $q \neq 0$ the singularity of $\varphi$ at $\rho=0$ gives a delta function
contribution to the scalar curvature at the origin of the Euclidean metric 
obtained by replacing in (\ref{flatwo}), $\eta \rightarrow i \theta$, {\it i.e.},
\begin{equation}
\sq \varphi = 4\pi q \frac{\delta (\rho)}{\rho}\,,
\label{delsingtwo}
\end{equation}
so that (\ref{philog}) should properly be considered a solution of
(\ref{auxeomtwo}) only in the distributional sense. If the point at 
$\rho^{-1} = 0$ is included in the manifold then there is also a
delta function contribution to (\ref{delsingtwo}) at infinity. In the Euclidean 
signature metric, the operator $\sq$ is the usual two dimensional Laplacian 
and (\ref{delsingtwo}) is Laplace's equation for the electrostatic potential 
with a point charge of magnitude $-q$ at the origin, and $+q$ at infinity. 
A non-zero value of $q$ corresponds to a topological defect on the Minkowski 
light cone, with the metric $e^{-\varphi}\, ds^2$, conformal to the original 
metric (\ref{flatwo}) possessing a conical singularity at $\rho = 0$ for 
generic $q \neq 0$.

Substituting (\ref{philog}) into the stress tensor (\ref{anomTtwo}),
we find in the $(\eta, \rho)$ coordinates,
\begin{equation}
T_a^{\ b} = \frac{Q^2}{4\pi} \frac{(2 - q) q}{\rho^2} \,
\left( \begin{array}{rc} -1 & 0 \\ 0 & 1 \end{array} \right)\,.
\label{TRindtwo}
\end{equation}
The case $q=0$ corresponds to the Minkowski vacuum with no singularity on 
the light cone, $t = \pm x$. However when $q = 1$, 
\begin{equation}
e^{-\varphi}\big\vert_{q=1}\, ds^2 = \rho_0^2\left(- d\eta^2 + 
\frac{d\rho^2}{\rho^2}\right) = \rho_0^2(-d\eta^2 + d\xi^2)\,,
\label{conftwo}
\end{equation}
with $\xi = \ln(\rho/\rho_0)$, is again flat. The vanishing Euler number of this
flat metric may be regarded as resulting from the cancellation of the $-1$ Euler 
charge at the origin from (\ref{auxeomtwo}), (\ref{Eulertwo}) and (\ref{delsingtwo})
with the $+1$ Euler number of a circular flat disc with boundary $\partial V$ at 
large but finite radius, as the boundary is taken to infinity in the original metric. 
In the conformal metric (\ref{conftwo}) in $(\eta, \xi)$ coordinates, for $q=1$ both 
singularities at $\rho =0$ and $\rho =\infty$ are removed to infinity and we again 
obtain a flat spacetime with no singularities and no boundaries. The original flat 
spacetime (\ref{flatwo}) has been conformally mapped to another flat spacetime with 
a different global topology. Indeed the Euclidean signature metric 
$d\theta^2 + d\xi^2$ with $\theta$ $2\pi$-periodic is the metric of $R \times S^1$, 
with punctures at $\xi = \pm \infty$. The Euler charge of $\pm 1$ unit at 
$\xi = \pm \infty$ is also a consequence of general theorems relating the Euler 
number to the number of  fixed points of the Killing field 
$K=\frac{\partial}{\partial \theta}$, where the metric (\ref{flatwo}) 
(or its inverse) becomes singular in the original $(\rho, \eta)$ 
coordinates \cite{Kill}.

When $q = 2$ the metric, $e^{-\varphi}\, ds^2$ becomes equivalent to the
original flat metric (\ref{flatwo}) by the change of variables
$\rho \rightarrow \rho_0^2/\rho$ which turns the spacetime
inside out, exchanging the coordinate singularities at $\rho =0$ 
and $\rho = \infty$. Since the new metric is equivalent to the original
metric up to a (singular) coordinate transformation, the stress tensor 
(\ref{TRindtwo}) again vanishes for $q = 2$. The invariance of the physical
stress tensor (\ref{TRindtwo}) under the transformation $q \rightarrow 2-q$
is evidence for a topological two-fold degenerate vacuum in two dimensional
gravity. 

For a single free scalar field $Q^2 = -1/6$ and at the value of $q=1$, 
(\ref{TRindtwo}) gives
\begin{equation}
T_a^{\ b}\Big\vert_{FR} = -\frac{1}{24\pi\rho^2} \,
\left( \begin{array}{rc} -1 & 0 \\ 0 & 1 \end{array} \right)\,,
\label{FulRintwo}
\end{equation}
in the $(\eta, \rho)$ coordinates. This is exactly the expectation value 
of the stress tensor of a quantum scalar field in the boost invariant 
Fulling-Rindler state \cite{FRtwo}. The preceding discussion shows that the 
divergence of $T_a^{\ b}$ in the Fulling-Rindler state is associated with a unit
topological defect on the Minkowski light cone, contained in the auxiliary 
potential $\varphi$, corresponding to a puncture at $\rho =0$ (equivalently, 
$\rho = \infty$) in the analytically continued Euclidean signature
metric. A singular conformal transformation (\ref{philog}) with $q=1$ is 
required to transform the Minkowski vacuum to the Fulling-Rindler state. 
Whereas the Hamiltonian defined with respect to the global Minkowski 
time coordinate $t$ is bounded from below and possesses a well-defined vacuum 
state, the boost generator corresponding to the Killing field
$\partial_\eta$ changes sign and is unbounded from below. These global
properties distinguishing the Fulling-Rindler state from the Minkowski vacuum 
are reflected in the topological defect (\ref{delsingtwo}) and
divergence of the stress-energy tensor (\ref{FulRintwo}) on the light cone.

A second example of macroscopic effects of topology of the trace
anomaly is provided by two-dimensional static metrics of the form,
\begin{equation}
ds^2 = -f dt^2 + \frac{dr^2}{f} = f (-dt^2 + dr^{*\,2}) \,,
\label{Stwo}
\end{equation}
where $f=f(r)$ and $dr^* = dr/f$. This class of metrics include
the two-dimensional Schwarzschild metric, with $f(r) = 1 - 2M/r$
and de Sitter metric in static coordinates, with $f(r) = 1 - H^2r^2$.
The Killing field $K = \partial_t$ is timelike for $f > 0$, and the scalar 
curvature corresponding to (\ref{Stwo}) is $R= - f''$. Hence in this case a
particular inhomogeneous solution to (\ref{auxeomtwo}) is 
$\varphi = \ln (-K^aK_a)^{\frac{1}{2}} = \ln f$. The conformally 
transformed line element $e^{-\varphi} ds^2 = -dt^2 + dr^{*\,2}$ is then the
flat optical metric \cite{opt}. 

The components of the stress tensor due to the anomaly are given by
\begin{subequations}
\begin{eqnarray}
T_{r^*}^{\ r^*} &=& \frac{N}{24\pi f} \left\{\frac{\varphi_{,r^*r^*}}{f} 
- \frac{f'}{2f} \varphi_{,r^*} + 
\frac{1}{4f} \left(\varphi_{,r^*}^2 + \dot\varphi^2\right)
+ R\right\}\,\\
T_t^{\ t} &=& \frac{N}{24\pi} \left\{-\frac{\ddot\varphi}{f}
+ \frac{f'}{2f} \varphi_{,r^*} 
- \frac{1}{4} \left(\varphi_{,r^*}^2 + \dot\varphi^2\right)
+ R\right\}\,,\qquad {\rm and}\\
T_t^{\ r^*} &=& \frac{N}{24\pi f} \left\{\dot \varphi_{,r^*}
- \frac{f'}{2} \dot\varphi 
+ \frac{1}{2} \dot\varphi \varphi_{,r^*} \right\}\,.
\end{eqnarray}
\end{subequations}
The off-diagonal flux component is generally time dependent unless 
we restrict ourselves to at most linear functions of time. With this 
restriction, the general solution of (\ref{auxeomtwo}) leading to a stationary 
stress tensor in the Schwarzschild case is:
\begin{equation}
\varphi = c_0 + \frac{q}{2M}\, r^* + \frac{p}{2M}\,t + \ln f\,,
\label{solphi}
\end{equation}
where $c_0$, $q$ and $p$ are constants. As in the Rindler case this is
a solution to (\ref{auxeomtwo}) in the distributional sense,
since $\sq \varphi$ contains delta function singularities.
Indeed with $p=0$ the strength of the delta function at the horizon,
mapped to a point on the Euclidean section $t\rightarrow i\tau$ for real $\tau$,
is $4\pi(q+1)$. For $q=-1$ this horizon singularity is cancelled, although
that value leads to a divergence in (\ref{solphi}) as $r\rightarrow \infty$.

The stress-energy tensor for $N$ scalar (or fermion) fields in the Schwarzschild 
case with $f= 1 - 2M/r$ is 
\begin{subequations}
\begin{eqnarray}
&&T_t^{\ t} = \frac{N}{24\pi}\left\{- \frac{1}{4f}
\left( \frac{p^2 + q^2}{4M^2} - \frac{4M^2}{r^4} \right)
+ \frac{4M}{r^3} \right\}\,,\\
&&T_t^{\ r^*} = \frac{N}{192\pi M^2} \frac{pq}{f}\,,\\
&&T_{r^*}^{\ r^*} = \frac{N}{96\pi f}
\left( \frac{p^2 + q^2}{4M^2} - \frac{4M^2}{r^4} \right)\,.
\end{eqnarray}
\label{TStwo}
\end{subequations}
\vskip-.3cm
\noindent This general form of the stress tensor for stationary states
in the two-dimensional Schwarzschild metric was obtained in ref. \cite{ChrFul},
from considerations of conservation, stationarity, and the trace anomaly. The
identification of the arbitrary constants $K$ and $Q$ in Eqs. (3.4) of that 
work with $-pq/192\pi$ and $[(p-q)^2 - 1]/384\pi$ respectively of the present 
article shows that the stress tensor obtained from the anomaly, (\ref{anomTtwo}) 
with (\ref{solphi}) coincides with that obtained in \cite{ChrFul}.

The stress tensor (\ref{TStwo}) is generally singular at $r=2M$, diverging as 
$f^{-1}$ as $f \rightarrow 0$. The condition for finiteness on the future horizon is 
\begin{equation}
\Big\vert \frac{1}{f} (T_t^{\ t} - T_{r^*}^{\ r^*} + 2 T_t^{\ r^*})\Big\vert
< \infty \qquad {\rm as} \qquad r\rightarrow 2M\,.
\label{finT}
\end{equation}
Substituting (\ref{TStwo}) into this condition and expanding about
$r=2M$ gives the condition,
\begin{equation}
(q-p)^2 = 1\,, \qquad {\rm (finiteness\ on\ future\ horizon)}
\label{finite}
\end{equation}
Finiteness on the past horizon would imply $(q+p)^2 = 1$. Thus, finiteness 
on both horizons implies , $p^2 + q^2 = 1$, $pq=0$ and zero flux. Since 
(\ref{finite}) cannot be satisfied with both $q$ and $p$ equal to zero, 
the solution (\ref{solphi}) must diverge as either $r^*$ or $t$ goes to 
infinity. Note that this is completely unlike the previous Fulling-Rindler 
case in which a finite (in fact, vanishing) stress tensor on the horizon 
is obtained with $\varphi = 0$,which has both zero charge on the light cone, 
{\it and} no divergence at infinity.

A time independent state with zero flux satisfying (\ref{finite})
is the two-dimensional Hartle-Hawking state with either $p$ or $q$
equal to zero, and the other eqial to $\pm 1$. 
The stress tensor for this state is given by
\begin{subequations}
\begin{eqnarray}
&&T_t^{\ t}\Big\vert_{HH} = \frac{N}{24\pi}\left\{\frac{1}{4f}
\left(\frac{4M^2}{r^4}  - \frac{1}{4M^2}\right)
+ \frac{4M}{r^3} \right\}\,,\\
&&T_t^{\ r^*}\Big\vert_{HH} = 0\,,\\
&&T_{r^*}^{\ r^*}\Big\vert_{HH} =  \frac{N}{96\pi f}
\left( \frac{1}{4M^2} - \frac{4M^2}{r^4} \right)\,.
\end{eqnarray}
\label{TSScht}
\end{subequations}
\vskip-.3cm
\noindent The degeneracy of the stress tensor for two different 
values of $p$ or $q$ is similar to the behavior
obtained in the Rindler case, (\ref{TRindtwo}).
As the horizon $r \rightarrow 2M$ is approached,
\begin{equation}
T_a^{\ b}\Big\vert_{HH} = \frac{\pi N T_H^2}{6} \,
\left( \begin{array}{rc} -1 & 0 \\ 0 & 1 \end{array} \right)\,,
\qquad r \rightarrow 2M\,,
\label{HHtwo}
\end{equation}
where $T_H = (8\pi M)^{-1}$ is the Hawking temperature. Thus the
stress tensor approaches that of a two dimensional perfect gas
at $T=T_H$ near the horizon in the Hartle-Hawking state. The
same value is also obtained in the limit $r\rightarrow \infty$,
showing that this state has infinite total energy. 

Another state obeying the finiteness condition (\ref{finT}) on the
future horizon is the Unruh state $p = -q = 1/2$. Its time inverse, 
given by $p = q =-1/2$ is finite on the past horizon. These states
have non-zero flux $T_t^{\ r^*}$ either outwardly or inwardly
directed from the horizon extending to infinity. The non-zero $q$
in these states reflects singularities, {\it i.e.} sources or sinks 
of the current (\ref{Noetwo}) at both $r=0$ and $r=\infty$.

Inspection of (\ref{TStwo}) shows that the only state which has vanishing stress 
tensor as $r\rightarrow \infty$ is the Boulware state with $p = q =0$, which 
does not satisfy the finiteness condition (\ref{finT}) at the horizon. Unlike
the previous Fulling-Rindler example in which finiteness at both the horizon
and infinty is achieved by one choice of $\varphi =0$, in the Schwarzschild
metric (\ref{Stwo}) because of its different topology, one is forced to choose 
between finiteness on the horizon and falloff at infinity. In the Boulware state 
with $p=q=0$,
\begin{equation}
T_a^{\ b}\Big\vert_{B} \rightarrow -\frac{\pi}{6} \frac{T_H^2}{f}\,
\left( \begin{array}{rc} -1 & 0 \\ 0 & 1 \end{array} \right)\,,
\qquad r \rightarrow 2M\,,
\label{Boultwo}
\end{equation}
as $r \rightarrow 2M$ and $f\rightarrow 0$, corresponding to the
{\it negative} of the stress tensor of a thermal distribution 
of massless particles at the blue-shifted local temperature 
$T_{loc} = T_H/\sqrt f$. 

Comparing (\ref{Boultwo}) with (\ref{FulRintwo}) we observe that the behavior 
of the stress tensor in the Boulware state in two dimensional Schwarzschild 
spacetime is {\it locally} the same as that of the Fulling-Rindler state 
in flat spacetime near the corresponding horizons, and indeed with the 
correspondence, $\rho \leftrightarrow 4M\sqrt f$, the behavior near the 
horizons in the two cases may be mapped onto each other. However the two 
situations are quite different {\it globally}. The Euler number of flat space is 
zero, and no singular behavior of $\varphi$ is required on either the light cone 
$t= \pm x$ or infinity. The Fulling-Rindler state can be obtained only by 
requiring a non-zero topological charge $q=1$ in the auxiliary field, which leads 
to $\varphi$ of (\ref{philog}) diverging at both large and small $\rho$.

On the other hand the scalar curvature $R=-f''$ is non-zero in the Schwarzschild 
case and the Euler number of the corresponding non-singular Euclidean signature 
metric is $+1$. This non-trivial topology of the background spacetime requires 
an inhomogeneous solution to (\ref{auxeomtwo}), such as $\varphi = \ln f$ in 
(\ref{solphi}). The divergence on the horizon of this particular solution to the 
inhomogeneous equation can be cancelled by homogeneous solutions with
non-zero $p$ and/or $q= -1$ satisfying (\ref{finite}), but only at the price of a
non-zero energy density, and non-trivial topological charge at infinity. 
Conversely the solution of (\ref{solphi}) with $c_0=p=q=0$ which vanishes at infinity 
and produces a transformed metric $e^{-\varphi} ds^2$ which is flat there, necessarily 
has a diverging stress tensor on the horizon. Thus while particular
solutions with particular boundary conditions for the auxiliary field can be chosen,
the impossibility of producing a solution of (\ref{solphi}) which is well-behaved 
at {\it both} the horizon and infinity is a global property of the Schwarzschild
spacetime (\ref{Stwo}), which is quite distinct from flat spacetime in Rindler
coordinates (\ref{flatwo}). The stress tensor of the anomaly and the auxiliary
scalar $\varphi$ captures these global effects of the macroscopic quantum
state in terms of the classical solutions of (\ref{auxeomtwo}).

Similar conclusions obtain in the two dimensional de Sitter spacetime
(\ref{Stwo}) with $f= 1-H^2r^2$. The general solution of (\ref{auxeomtwo})
leading to a stationary stress tensor is
\begin{equation}
\varphi = c_0 + 2 q H r^* + 2 p H t + \ln f\,. 
\label{solphids}
\end{equation}
The components of the stress tensor in the general stationary state labelled
by $(p,q)$ are
\begin{subequations}
\begin{eqnarray}
&&T_t^{\ t} = \frac{NH^2}{24\pi}\left\{ - \frac{1}{f}
\left( p^2 + q^2 - H^2r^2 \right) + 2 \right\}\,,\\
&&T_{r^*}^{\ r^*} =  \frac{NH^2}{24\pi f}
\left( p^2 + q^2  - H^2r^2 \right)\,.\\
&&T_t^{\ r^*} = \frac{NH^2}{12\pi}\,\frac{pq}{f}\,.
\end{eqnarray}
\label{TStwods}
\end{subequations}
\vskip-.3cm
\noindent The generic stress tensor again diverges on the horizon where 
$f=1-H^2r^2 =0$, despite the finiteness of the scalar curvature 
$R=-f'' = 2H^2$ there. The condition that the stress tensor be finite on the 
future horizon is again $(q-p)^2 = 1$. The divergence of the stress tensor can 
be cancelled only by the addition of a specific homogeneous solution to 
(\ref{auxeomtwo}). States which are regular on both the past and future 
horizon have $p^2 + q^2 = 1$, $pq=0$ and zero flux. These conditions give the 
Bunch-Davies state with the non-vanishing de Sitter invariant stress tensor, 
$T_t^{\ t} = T_{r^*}^{\ r^*} = NH^2/24\pi$. The analog of the Boulware state 
with $p=q=0$ diverges as $f\rightarrow 0$ exactly as in (\ref{Boultwo}) with 
$T_H = H/2\pi$. 

In all the two dimensional examples considered, the parameter 
$q$ may be viewed as an effective topological charge carried by the 
auxiliary field of the trace anomaly. The classical solutions of this 
order parameter field generally depend on global coordinate invariant 
quantities, such as $\ln (-K^aK_a)$ which diverges on a horizon
where a timelike Killing field $K^a$ becomes null, notwithstanding
the finiteness of the local curvature there. Thus the scalar auxiliary
field provides a coordinate invariant characterization of the 
local divergences of the stress-energy on the horizon, in macroscopic 
quantum states defined globally in the spacetime. 

\section{Effective Action and Stress Tensor in Four Dimensions}

The low energy effective action for gravity in four dimensions 
contains first of all, the local terms constructed from the
Riemann curvature tensor and its derivatives and contractions up to
and including dimension four. This includes the usual Einstein-Hilbert
action of general relativity,
\begin{equation}
S_{EH}[g] = \frac{1}{16\pi G} \int\, d^4x\,\sqrt{-g}\, (R-2\Lambda)  
\label{clfour}
\end{equation}
as well as the spacetime integrals of the fourth order curvature
invariants,
\begin{equation}
S_{local}^{(4)}[g] = \frac{1}{2} \int\, \sqrt{-g}\, (\alpha C_{abcd}C^{abcd}
+ \beta R^2)\, d^4x\,,
\label{R2}
\end{equation}
with arbitrary dimensionless coefficients $\alpha$ and $\beta$. There are two 
additional fourth order invariants, namely $E= ^*\hskip-.2cm R_{abcd}\,
^*\hskip-.1cm R^{abcd}$ and $\sq R$, which could be added to (\ref{R2}) as well, 
but as they are total derivatives yielding only a surface term and no local 
variation, we omit them. All the possible local terms in the effective action 
may be written as the sum,
\begin{equation}
S_{local}[g] = {1\over 16\pi G}\int\,d^4x\,\sqrt{-g}\,(R-2\Lambda) +
S_{local}^{(4)} + \sum_{n=3}^{\infty} S_{local}^{(2n)}\,.
\label{locsum}
\end{equation}
with the terms in the sum with $n\ge 3$ composed of integrals of local curvature 
invariants with dimension $2n \ge 6$, and suppressed by $M^{-2n + 4}$ at energies 
$E \ll M$. Here $M$ is the ultraviolet cutoff scale of the low energy effective 
theory which we may take to be of order $M_{Pl}$. The higher derivative terms 
with $n \ge 3$ are irrelevant operators in the infrared, scaling with negative 
powers of $e^{\sigma_0}$ under global rescalings of the metric, as in 
(\ref{metconf}), and may be neglected at macroscopic distance scales. On the 
other hand the two terms in the Einstein-Hilbert action $n= 0, 1$ scale as 
$e^{4\sigma_0}$ or $e^{2\sigma_0}$ respectively, and are clearly relevant 
in the infrared. The fourth order terms in (\ref{R2}) are neutral under 
such global rescalings. 

The exact quantum effective action also contains non-local terms in general. 
All possible terms in the effective action (local or not) can be classified 
according to how they respond to global Weyl rescalings of the metric. If the 
non-local terms are non-invariant under global rescalings, then they  
scale either positively or negatively under 
$g_{ab} \rightarrow e^{2\sigma_0} g_{ab}$. If $m^{-1}$ is some fixed 
length scale associated with the non-locality, arising for example 
by the integrating out of fluctuations of fields with mass $m$, then 
at much larger macroscopic distances ($mL \gg 1$) the non-local terms 
in the effective action become approximately local. The terms which 
scale with positive powers of $e^{\sigma_0}$ are constrained by general 
covariance to be of the same form as the $n=0,1$ Einstein-Hilbert terms 
in $S_{local}$, (\ref{clfour}). Terms which scale negatively with 
$e^{\sigma_0}$ become negligibly small as $mL \gg 1$ and are infrared 
irrelevant at macroscopic distances. This is the expected decoupling of 
short distance degrees of freedom in an effective field theory description, 
which are verified in detailed calculations of loops in massive field 
theories in curved space. The only possibility for contributions 
to the effective field theory of gravity at macroscopic distances,
which are not contained in the local expansion of 
(\ref{locsum}) arise from fluctuations not associated with any finite 
length scale, {\it i.e.} $m=0$. These are the non-local contributions to 
the low energy EFT which include those associated with the anomaly.

Classical fields satisfying wave equations with zero mass, which are
invariant under conformal transformations of the spacetime metric, 
$g_{ab} \rightarrow e^{2\sigma} g_{ab}$ have stress tensors with zero 
classical trace, $T_a^{\ a} = 0$. Because the
corresponding quantum theory requires a UV regulator, classical conformal 
invariance cannot be maintained at the quantum level. The
trace of the stress tensor is generally non-zero when $\hbar \ne 0$, and 
any UV regulator which preserves the covariant conservation 
of $T_a^{\ b}$ (a necessary requirement of any theory respecting general
coordinate invariance) yields an expectation value of the quantum 
stress tensor with the non-zero trace \cite{anom,BirDav}, 
\begin{equation}
\langle T_a^{\ a} \rangle
= b F + b' \left(E - \frac{2}{3}\sq R\right) + b'' \sq R + \sum_i \beta_iH_i\,,
\label{tranom}
\end{equation}
in a general four dimensional curved spacetime. This is the four
dimensional analog of (\ref{trtwo}) in two dimensions.
In Eq. (\ref{tranom}) we employ the notation,
\begin{subequations}
\begin{eqnarray}
&&E \equiv ^*\hskip-.2cmR_{abcd}\,^*\hskip-.1cm R^{abcd} = 
R_{abcd}R^{abcd}-4R_{ab}R^{ab} + R^2 
\,,\qquad {\rm and} \label{EFdef}\\
&&F \equiv C_{abcd}C^{abcd} = R_{abcd}R^{abcd}
-2 R_{ab}R^{ab}  + {R^2\over 3}\,.
\end{eqnarray}
\end{subequations}

\noindent  with $R_{abcd}$ the Riemann curvature tensor, 
$^*\hskip-.2cmR_{abcd}= \varepsilon_{abef}R^{ef}_{\ \ cd}/2$ its dual, 
and $C_{abcd}$ the Weyl conformal tensor. The coefficients $b$, $b'$ 
and $b''$ are dimensionless parameters proportional to $\hbar$. These terms 
appear in the trace of stress tensor of any massless field, provided only that it is 
regularized and defined in such a way as to preserve general coordinate invariance 
and the covariant conservation law, $\nabla_a \langle T^a_{\ b}\rangle = 0$, which
we assume if the theory is to preserve general covariance. Additional terms denoted
by the sum $\sum_i \beta_i H_i$ in (\ref{tranom}) may also appear in the general
form of the trace anomaly, if the massless field in question couples to additional
long range gauge fields. Thus in the case of massless fermions coupled to
a background gauge field, the invariant $H =$tr($F_{ab}F^{ab}$) will appear in 
(\ref{tranom}) with a coefficient $\beta$ determined by the beta function of
the relevant gauge coupling. We shall concentrate on the universal gravitational
terms in the anomaly in most of what follows, returning to consider the effect of 
the $\sum_i \beta_i H_i$ terms at the end of this section.

The form of (\ref{tranom}) and coefficients $b$ and $b'$ do not depend on the state 
in which the expectation value of the stress tensor is computed. Instead they determined 
only by the number of massless fields \cite{BirDav}, 
\begin{subequations}
\begin{eqnarray}
b &=& \frac{1}{120 (4 \pi)^2}\, (N_S + 6 N_F + 12 N_V)\,,\\
b'&=& -\frac{1}{360 (4 \pi)^2}\, (N_S + 11 N_F + 62 N_V)\,,
\end{eqnarray}
\label{bbprime}
\end{subequations}
\vskip -.3cm
\noindent with $(N_S, N_F, N_V)$ the number of fields of spin $(0, \frac{1}{2}, 1)$
respectively and we have taken $\hbar = 1$.

Three local fourth order curvature invariants $E, F$ and $\sq R$ appear in the 
trace of the stress tensor, but only the first two ($b$ and $b'$) terms of 
(\ref{tranom}) cannot be derived from a local effective action of the metric
alone. If these terms could be derived from a local gravitational action we could 
simply make the necessary finite redefinition of the corresponding local counterterms 
to remove them from the trace, in which case the trace would no longer be non-zero 
or anomalous. This redefinition of a local counterterm (namely, the $R^2$ term in 
the effective action) is possible only with respect to the third $b''$ coefficient 
in (\ref{tranom}), which is therefore regularization dependent and not part of 
the true anomaly. Only the non-local effective action corresponding to the $b$ 
and $b'$ terms in (\ref{tranom}) lead to the possibility of effects that extend 
over arbitrarily large, macroscopic distances, unsuppressed by any ultraviolet 
cutoff scale. The distinction of the two kinds of terms in the effective
action (local or not) is emphasized in the cohomological approach
to the trace anomaly \cite{MazMot}.

The number of massless fields of each spin, $N_S, N_F, N_V$ is a property of the 
low energy effective description of matter, having no direct connection with physics at the 
ultrashort Planck scale. Indeed massless fields fluctuate at all distance scales 
and do not decouple in the far infrared. As in the case of the chiral anomaly with
massless quarks, the $b$ and $b'$ terms in the trace anomaly were calculated
originally by techniques usually associated with UV regularization (such as
dimensional regularization, point splitting or heat kernel techniques) \cite{BirDav}. 
However just as in the case of the chiral anomaly in QCD, (\ref{tranom}) and 
(\ref{bbprime}) can have significant effects in the far infrared as well.

To find the WZ effective action corresponding to the $b$ and $b'$ terms in
(\ref{tranom}), introduce as in two dimensions the conformal parameterization 
(\ref{confac}), and compute
\begin{subequations}
\begin{eqnarray}
\sqrt{-g}\,F &=& \sqrt{\bar g}\,\bar F\,\\
\sqrt{-g}\,\left(E - {2\over 3}\sq R\right) &=& \sqrt{-g}\,
\left(\overline E - {2\over 3}\sqb\overline R\right) + 4\,\sqrt{\bar g}\,
\bar\Delta_4\,\sigma\,,
\label{Esig}
\end{eqnarray}
\label{FEsig}
\end{subequations}
\vskip-.3cm
\noindent whose $\sigma$ dependence is no more than linear. The 
fourth order differential operator appearing in this expression is \cite{Rie,AM,MazMot}
\begin{equation}
\Delta_4 \equiv \sq^2 + 2 R^{ab}\nabla_a\nabla_b - {2\over 3} R \sq + 
{1\over 3} (\nabla^a R)\nabla_a \,,
\label{Deldef}
\end{equation}
which is the unique fourth order scalar operator that is conformally covariant, 
{\it viz.}
\begin{equation}
\sqrt{-g} \Delta_4 = \sqrt{-\bar g} \bar \Delta_4 \,,
\label{invfour}
\end{equation}
for arbitrary smooth $\sigma(x)$ in four (and only four) dimensions. 
Thus multiplying (\ref{tranom}) by $\sqrt{-g}$ and recognizing that the
result is the $\sigma$ variation of an effective action $\Gamma_{WZ}$, we
find immediately that up to terms independent of $\sigma$, this
effective action is
\begin{equation}
\Gamma_{WZ}[\bar g;\sigma] = b  \int\,d^4x\,\sqrt{-\bar g}\, \bar F\,\sigma
+ b' \int\,d^4x\,\sqrt{-\bar g}\,\left\{\left(\bar E - {2\over 3}
\sqb \bar R\right)\sigma + 2\,\sigma\bar\Delta_4\sigma\right\}\,.
\label{WZfour}
\end{equation}
This Wess-Zumino action is a one-form representative of the non-trivial 
cohomology of the local Weyl group and satisfies (\ref{WZcob}) just as
its two dimensional analog $\Gamma_{WZ}^{(2)}$ does. Explicitly,
by solving (\ref{Esig}) formally for $\sigma$ and substituting
the result in (\ref{WZfour}) we obtain
\begin{equation}
\Gamma_{WZ}[\bar g;\sigma] = \Delta_\sigma \circ S_{anom}[\bar g] \equiv
S_{anom}[g=e^{2\sigma}\bar g] - S_{anom}[\bar g],
\label{Weylshift}
\end{equation}
with
\begin{equation}
 S_{anom}[g] =  {1\over 8}\int d^4x\sqrt{-g}\int d^4x'\sqrt{-g'}
\left(E - {2\over 3} \sq R\right)_x\Delta_4^{-1} (x,x')\left[ 2bF + b'
\left(E - {2\over 3} \sq R\right)\right]_{x'}\hskip-.1cm 
\label{nonl}
\end{equation}
and $\Delta_4^{-1}(x,x')$ denoting the Green's function inverse of the fourth 
order differential operator defined by (\ref{Deldef}). From the foregoing
construction it is clear that if there are additional Weyl invariant
terms in the anomaly (\ref{tranom}) they should be included in the
$S_{anom}$ by making the replacement $bF \rightarrow bF + \sum_i\beta_i H_i$
in the last square bracket of Eq. (\ref{nonl}).

Because of (\ref{Weylshift}) the anomalous effective action is clearly fixed only 
up to an arbitrary (local or non-local) Weyl invariant functional $S_{inv}[g]$ of 
the metric, obeying
\begin{equation}
\Delta_{\sigma}\circ S_{inv}[g] = S_{inv}[e^{2\sigma} g] - S_{inv}[g] = 0\,.
\label{Sinv}
\end{equation}
Hence the general non-local form of the {\it exact} effective action of
gravity must be of the form, $S_{anom} + S_{inv}$, and behavior under the local 
Weyl group enables us to classify all terms in the exact effective action 
into three parts, {\it viz.}
\begin{equation}
S_{exact} = S_{local} + S_{anom} + S_{inv}\,,
\label{Sexact}
\end{equation}
given by (\ref{locsum}), (\ref{nonl}) and (\ref{Sinv}) respectively. As in two 
dimensions, the anomalous term (\ref{WZfour}) or (\ref{nonl}) scales linearly 
with $\sigma_0$, {\it i.e.} logarithmically with distance, and is therefore
{\it marginally relevant} in the infrared with respect to the usual classical 
flat space fixed point \cite{MazMot}. The fluctuations generated by $S_{anom}$
also define a non-perturbative Gaussian infrared fixed point, with conformal
field theory anomalous dimensions analogous to the two dimensional case \cite{AM}.
This is possible only because new low energy degrees of freedom are contained in
$S_{anom}$ which can fluctuate independently of the local metric degrees of 
freedom in $S_{EH}$. Thus the effective action of the anomaly $S_{anom}$ should be 
retained in the EFT of low energy gravity, which is specified then by the first 
two strictly relevant local terms of the classical Einstein-Hilbert action 
(\ref{clfour}), and the logarithmic, marginally relevant $S_{anom}$, {\it i.e.}
\begin{equation}
S_{eff}[g] = S_{EH}[g] + S_{anom}[g]
\label{Seff}
\end{equation}
contains all the infrared relevant terms in low energy gravity for 
$E \ll M_{Pl}$. 

The low energy (Wilson) effective action (\ref{Seff}), in which infrared 
irrelevant terms are systematically neglected in the renormalization
group program of critical phenomena is to be contrasted with the exact 
(field theoretic) effective action of (\ref{Sexact}), in which the effects 
of all scales are included in principle, at least in the approximation 
in which spacetime can be treated as a continuous manifold. Ordinarily, 
{\it i.e.} absent anomalies, the Wilson effective action should contain 
only local infrared relevant terms consistent with symmetry \cite{Wil}. 
However, like the anomalous effective action generated by the chiral 
anomaly in QCD, the non-local $S_{anom}$ must be included in the low 
energy EFT to account for the anomalous Ward identities, even in the 
zero momentum limit; and indeed logarithmic scaling with distance 
indicates that $S_{anom}$ is an infrared relevant term. Also even if 
no massless matter fields are assumed, the quantum fluctuations of 
the metric itself will generate a term of the same form as $S_{anom}$
\cite{AMMc}. Since all the infrared relevant terms are now included 
in $S_{eff}$, it contains all the low energy degrees of freedom of 
gravity, including also those massless scalar degrees of freedom in 
the local auxiliary field description which we exhibit below, and 
which are those responsible for the non-perturbative infrared fixed 
point with vanishing $\lambda$ found in \cite{AM}. In contrast, the 
additional terms in $S_{exact}$, being either marginally or strictly 
irrelevant in the infrared limit, contain non-universal and state 
dependent contributions which are not required for a consistent EFT 
expansion of the gravitational effective action in powers of $E/M_{Pl}$. 
Since anomalies are the only way known that the decoupling hypothesis can 
fail, the neglected terms in $S_{inv}$ just as those in $S_{local}$ for 
$n\ge 2$ are also not expected to require the addition of any new degrees 
of freedom of the gravitational field which are relevant at low energies 
or macroscopic distances. 
 
The variations of the two ($b$ and $b'$) terms in $S_{anom}$ give rise to 
two new conserved tensors in the equations of low energy gravity. To find 
their explicit form and exhibit the new scalar degrees of freedom they 
contain, it is convenient as in the two dimensional case to rewrite the 
non-local action $S_{anom}$ in local form by introducing auxiliary fields. 
Two scalar auxiliary fields satisfying
\begin{subequations}
\begin{eqnarray}
&& \Delta_4\, \varphi = {1\over 2} \left(E - {2\over 3} \sq R\right)\,,
\label{auxvarphi}\\
&& \Delta_4\, \psi = {1\over 2} F\,,
\label{auxvarpsi}
\end{eqnarray}
\label{auxeom}
\end{subequations}
\vskip -.5cm
\noindent respectively may be introduced, corresponding to the two non-trivial
cocycles of the $b$ and $b'$ terms in the anomaly \cite{MazMot}. It is then easy 
to see that 
\begin{eqnarray} 
&& S_{anom}[g;\varphi,\psi] =  {b'\over 2}\,\int\,d^4x\,\sqrt{-g}\ 
\left\{ -\varphi \Delta_4 \varphi + \left(E - {2\over 3} \sq R\right) \varphi \right\}
\nonumber\\
&& \quad + {b\over 2} \,\int\,d^4x\,\sqrt{-g}\ \left\{ -2\varphi \Delta_4 \psi +
F \varphi + \left(E - {2\over 3} \sq R\right) \psi \right\}
\label{locaux}
\end{eqnarray}
is the desired local form of the anomalous action (\ref{nonl}) \cite{BFS}. 
Indeed the variation of (\ref{locaux}) with respect to the auxiliary fields
$\varphi$ and $\psi$ yields their Eqs. of motion (\ref{auxeom}), which
may be solved for $\varphi$ and $\psi$ by introducing the Green's function 
$\Delta_4^{-1}(x,x')$. Substituting this formal solution for the auxiliary fields 
into (\ref{locaux}) returns (\ref{nonl}). The local auxiliary field
form (\ref{locaux}) is the most useful and explicitly contains two new scalar 
fields satisfying the massless fourth order wave equations (\ref{auxeom})
with fourth order curvature invariants as sources. The freedom to add homogeneous 
solutions to $\varphi$ and $\psi$ corresponds to the freedom to define
different Green's functions inverses $\Delta_4^{-1}(x,x')$ in (\ref{nonl}).
The auxilliary scalar fields are new local massless degrees of freedom of four 
dimesnional gravity, not contained in the Einstein-Hilbert action.

As in the two dimensional case, the four dimensional anomalous action
has an invariance, in this case $\psi \rightarrow \psi + \psi_0$, with
the corresponding Noether current given by:
\begin{equation}
J^a = \nabla^a \sq \varphi + 2 \left(R^{ab} - \frac{R}{3} g^{ab} \right)
\nabla_b \varphi - \frac{1}{2} \Omega^a + \frac{1}{3} \nabla^a R\,,
\label{Noether}
\end{equation}
where $\Omega^a$ is the topological current whose divergence,
\begin{equation}
\nabla_a \Omega^a = E = R_{abcd}R^{abcd} -4R_{ab}R^{ab} + R^2 
\end{equation}
is the Euler-Gauss-Bonnet integrand in four dimensions. The current $J^a$
is conserved,
\begin{equation}
\nabla_a J^a = \Delta_4 \varphi - \frac{E}{2} +\frac{1}{3} \sq R = 0
\end{equation}
by (\ref{auxvarphi}), and the conserved charge may be taken to be
\begin{equation}
q = \frac{1}{16\pi^2} \int_{\Sigma} J^a\,d\Sigma_a
\label{charge}
\end{equation}
with $\Sigma$ a spacelike Cauchy surface on which initial data are specified or 
\begin{equation}
q = \frac{1}{16\pi^2} \oint_{\partial V} J^a\,d\Sigma_a
\label{topcharge}
\end{equation}
with $\partial V$ the boundary of a Euclidean four volume $V$. The
normalization of the Noether charge is chosen so that $q=1$
corresponds to a delta function source of unit strength in 
the four dimensional Euler number,
\begin{equation}
\chi = \frac{1}{32\pi^2}\int\,d^4x\sqrt{-g}\, E\,.
\end{equation}
The constant shift in the other auxiliary field, {\it i.e.} $\varphi \rightarrow
\varphi + \varphi_0$ generates a global rescaling of the metric, and is not a 
symmetry of the action (\ref{locaux}) unless $E- 2\sq R/3 =F =0$. It is 
worth remarking that this enhanced conformal symmetry applies to solutions
of the vacuum Einstein Eqs. if and only if the cosmological term $\Lambda =0$.
That the effective action of the anomaly may provide a mechanism to relax
the cosmological term to zero was first proposed in ref. \cite{AM}.

By using the definition of $\Delta_4$ and integrating by parts, 
we may express the anomalous action also in the form,
\begin{equation} 
S_{anom} = b' S^{(E)}_{anom} + b S^{(F)}_{anom}\,,
\label{allanom}
\end{equation}
with
\begin{eqnarray}
&& S^{(E)}_{anom} \equiv {1\over 2} \int\,d^4x\,\sqrt{-g}\ \left\{
-\left(\sq \varphi\right)^2 + 2\left(R^{ab} - {R\over 3}g^{ab}\right)(\nabla_a \varphi)
(\nabla_b \varphi) + \left(E - {2\over 3} \sq R\right) \varphi\right\}\,;\nonumber\\
&& S^{(F)}_{anom} \equiv \,\int\,d^4x\,\sqrt{-g}\ \left\{ -\left(\sq \varphi\right)
\left(\sq \psi\right) + 2\left(R^{ab} - {R\over 3}g^{ab}\right)(\nabla_a \varphi)
(\nabla_b \psi)\right.\nonumber\\
&& \qquad\qquad + \left.{1\over 2} F \varphi + 
{1\over 2} \left(E - {2\over 3} \sq R\right) \psi \right\}
\label{SEF}
\end{eqnarray}
We note that the $S^{(E)}$ and $S^{(F)}$ terms are very similar. Each is
composed of terms at most quadratic and linear in the auxiliary fields
$\varphi$ and $\psi$. Defining the general quadratic action,
\begin{equation}
A[g;\varphi,\psi] \equiv \int\,d^4x\,\sqrt{-g}\ \left\{ -\left(\sq \varphi\right)
\left(\sq \psi\right) + 2\left(R^{ab} - {R\over 3}g^{ab}\right)(\nabla_a \varphi)
(\nabla_b \psi)\right\}
\label{quadr}
\end{equation}
and the two linear actions,
\begin{eqnarray}
&& B[g; \varphi] \equiv {1\over 2} \int\,d^4x\,\sqrt{-g}\ 
\left(E - {2\over 3} \sq R\right) \varphi \,;\\
&& C[g; \varphi] \equiv {1\over 2} \int\,d^4x\,\sqrt{-g}\ F \varphi\,,
\label{linr}
\end{eqnarray}
as functionals of the metric and the two auxiliary scalar fields 
$\varphi$ and $\psi$, we can express (\ref{SEF}) in the form,
\begin{subequations}
\begin{eqnarray}
&& S^{(E)}_{anom} = {1\over 2} \,A[g;\varphi,\varphi] + B[g; \varphi]\,;\\
&& S^{(F)}_{anom} = A[g;\varphi,\psi] + B[g; \psi] + C[g; \varphi]\,.
\end{eqnarray}
\label{SEFABC}
\end{subequations}

\noindent This implies that the covariantly conserved stress tensors derived 
from the $E$ and $F$ terms in the effective action, namely,
\begin{subequations}
\begin{eqnarray}
E_{ab} &\equiv & -{2\over \sqrt{-g}} {\delta S^{(E)}_{anom}\over \delta
g^{ab}}\,\\
F_{ab} &\equiv& -{2\over \sqrt{-g}} {\delta S^{(F)}_{anom}\over \delta
g^{ab}}\,.
\end{eqnarray}
\end{subequations}
\vskip-.3cm
\noindent can be obtained from one fundamental tensor quadratic in the scalar 
auxiliary fields,
\begin{eqnarray}
&&A_{ab} [g; \varphi, \psi] \equiv -{2\over \sqrt{-g}} {\delta
 A[g;\varphi,\psi]\over \delta g^{ab}} = 
-2\, (\nabla_{(a}\varphi) (\nabla_{b)} \sq \psi) 
-2\, (\nabla_{(a}\psi) (\nabla_{b)} \sq \varphi) \nonumber\\
&& 
\hspace{1cm} + 2\,\nabla^c \left[(\nabla_c \varphi)(\nabla_a\nabla_b\psi) 
 + (\nabla_c \psi)(\nabla_a\nabla_b\varphi)\right]
- {4\over 3}\, \nabla_a\nabla_b\left[(\nabla_c \varphi)
(\nabla^c\psi)\right]  
\nonumber\\ 
&&  
\hspace{.5cm} + {4\over 3}\,R_{ab}\, (\nabla_c \varphi)(\nabla^c \psi) 
- 4\, R^c_{\ (a}\left[(\nabla_{b)} \varphi) (\nabla_c \psi)
+ (\nabla_{b)} \psi) (\nabla_c \varphi)\right]
 + {4\over 3}\,R \,(\nabla_{(a} \varphi) (\nabla_{b)} \psi) \nonumber\\ 
&& 
\hspace{.5cm} + {1\over 3}\, g_{ab}\, \left\{-3\, (\sq\varphi)(\sq\psi)
+ \sq \left[(\nabla_c\varphi)(\nabla^c\psi)\right] 
+ 2\, \left( 3R^{cd} - R g^{cd} \right) (\nabla_c
\varphi)(\nabla_d \psi)\right\}\,,
\label{Aten}
\end{eqnarray}
which is traceless,
\begin{equation}
A_a^{\ a} [g; \varphi, \psi] = 0\,,
\end{equation}
for arbitrary $\varphi$ and $\psi$, and two fundamental tensors linear 
in the scalar auxiliary fields,
\begin{eqnarray}
&& B_{ab}[g; \varphi] \equiv -{2\over \sqrt{-g}} {\delta
 B[g;\varphi]\over \delta g^{ab}} =
- {2\over 3}\, \nabla_a\nabla_b \sq \varphi 
- 4\, C_{a\ b}^{\ c\ d}\, \nabla_c \nabla_d \varphi \nonumber\\
&& \hspace{.5cm} 
- 4\, R_{(a}^c \nabla_{b)} \nabla_c\varphi 
+ {8\over 3}\, R_{ab}\, \sq \varphi  
+ {4\over 3}\, R\, \nabla_a\nabla_b\varphi 
- {2\over 3} \left(\nabla_{(a}R\right) \nabla_{b)}\varphi \nonumber\\
&& 
\hspace{1cm}
+ {1\over 3}\, g_{ab}\, \left\{ 2\, \sq^2 \varphi 
+ 6\,R^{cd} \,\nabla_c\nabla_d\varphi 
- 4\, R\, \sq \varphi 
+ (\nabla^c R)\nabla_c\varphi\right\}\,,
\label{Bten}
\end{eqnarray}
and
\begin{equation}
C_{ab}[g; \varphi] \equiv -{2\over \sqrt{-g}} {\delta
 C[g;\varphi]\over \delta g^{ab}} =
- 4\, \nabla_c\nabla_d\left( C_{(a\ b)}^{\ \ c\ \ d}
\varphi \right)  - 2\, C_{a\ b}^{\ c\ d} R^{cd} \varphi\,.
\label{Cten}
\end{equation}
The latter of these is also traceless, whereas the $B_{ab}$ tensor has
a non-zero trace whose value depends on which auxiliary field
is substituted, {\it i.e.}
\begin{subequations}
\begin{eqnarray}
B_a^{\ a}[g;\varphi] &=& 2\Delta_4 \varphi = E - \frac{2}{3}\sq R\,\\
B_a^{\ a}[g;\psi] &=& 2\Delta_4 \psi = F \,.
\end{eqnarray}
\end{subequations}
From (\ref{allanom}), (\ref{SEF}), (\ref{SEFABC}), (\ref{Aten}), (\ref{Bten})
and (\ref{Cten}) we have
\begin{equation}
T_{ab}^{(anom)} \equiv -{2\over \sqrt{-g}} {\delta
 S_{anom}\over \delta g^{ab}} = b' E_{ab} + b F_{ab} 
\label{Tanom}
\end{equation}
with
\begin{subequations}
\begin{eqnarray}
&& E_{ab} = \frac{1}{2} A_{ab}[g;\varphi, \varphi] + B_{ab}[g; \varphi]
\qquad {\rm and}\\
&& F_{ab} = A_{ab}[g;\varphi, \psi] + B_{ab}[g; \psi] + C_{ab}[g; \varphi]\,.
\label{ABCF}
\end{eqnarray}
\end{subequations}
\vskip-.3cm
\noindent Explicitly,
\begin{eqnarray}
E_{ab} &=&-2\, (\nabla_{(a}\varphi) (\nabla_{b)} \sq \varphi) 
+ 2\,\nabla^c \left[(\nabla_c \varphi)(\nabla_a\nabla_b\varphi)\right]  
- {2\over 3}\, \nabla_a\nabla_b\left[(\nabla_c \varphi)
(\nabla^c\varphi)\right]\nonumber\\ 
&&  
+ {2\over 3}\,R_{ab}\, (\nabla_c \varphi)(\nabla^c \varphi)  
- 4\, R^c_{\ (a}(\nabla_{b)} \varphi) (\nabla_c \varphi)
+ {2\over 3}\,R \,(\nabla_a \varphi) (\nabla_b \varphi)\nonumber\\ 
&& + {1\over 6}\, g_{ab}\, \left\{-3\, (\sq\varphi)^2 
+ \sq \left[(\nabla_c\varphi)(\nabla^c\varphi)\right] 
+ 2\left( 3R^{cd} - R g^{cd} \right) (\nabla_c \varphi)(\nabla_d
\varphi)\right\}\nonumber\\ 
&&
\hspace{-1cm} - {2\over 3}\, \nabla_a\nabla_b \sq \varphi 
- 4\, C_{a\ b}^{\ c\ d}\, \nabla_c \nabla_d \varphi
- 4\, R_{(a}^c \nabla_{b)} \nabla_c\varphi 
+ {8\over 3}\, R_{ab}\, \sq \varphi  
+ {4\over 3}\, R\, \nabla_a\nabla_b\varphi \nonumber\\
&& 
- {2\over 3} \left(\nabla_{(a}R\right) \nabla_{b)}\varphi
+ {1\over 3}\, g_{ab}\, \left\{ 2\, \sq^2 \varphi 
+ 6\,R^{cd} \,\nabla_c\nabla_d\varphi 
- 4\, R\, \sq \varphi 
+ (\nabla^c R)\nabla_c\varphi\right\}\,,
\label{Eab}
\end{eqnarray}
and
\begin{eqnarray}
&& F_{ab} = -2\, (\nabla_{(a}\varphi) (\nabla_{b)} \sq \psi) 
-2\, (\nabla_{(a}\psi) (\nabla_{b)} \sq \varphi)
+ 2\,\nabla^c \left[(\nabla_c \varphi)(\nabla_a\nabla_b\psi) 
 + (\nabla_c \psi)(\nabla_a\nabla_b\varphi)\right]
\nonumber\\ 
&&  
\hspace{.5cm} - {4\over 3}\, \nabla_a\nabla_b\left[(\nabla_c \varphi)
(\nabla^c\psi)\right]
+ {4\over 3}\,R_{ab}\, (\nabla_c \varphi)(\nabla^c \psi) 
- 4\, R^c_{\ (a}\left[(\nabla_{b)} \varphi) (\nabla_c \psi)
+ (\nabla_{b)} \psi) (\nabla_c \varphi)\right]\nonumber\\ 
&& 
\hspace{1.5cm} + {4\over 3}\,R \,(\nabla_{(a} \varphi) (\nabla_{b)} \psi) 
+ {1\over 3}\, g_{ab}\, \Big\{-3\, (\sq\varphi)(\sq\psi)
+ \sq \left[(\nabla_c\varphi)(\nabla^c\psi)\right] 
\nonumber\\ 
&&
\hspace{1.5cm} \left. + 2\, \left( 3R^{cd} - R g^{cd} \right) (\nabla_c
\varphi)(\nabla_d \psi)\right\}- 4\, \nabla_c\nabla_d\left( C_{(a\ b)}^{\ \ c\ \ d}
\varphi \right)  - 2\, C_{a\ b}^{\ c\ d} R^{cd} \varphi \nonumber\\
&&
\hspace{.5cm} - {2\over 3}\, \nabla_a\nabla_b \sq \psi 
- 4\, C_{a\ b}^{\ c\ d}\, \nabla_c \nabla_d \psi
- 4\, R_{(a}^c (\nabla_{b)} \nabla_c\psi) 
+ {8\over 3}\, R_{ab}\, \sq \psi  
+ {4\over 3}\, R\, \nabla_a\nabla_b\psi \nonumber\\
&& 
\hspace{1cm} - {2\over 3} \left(\nabla_{(a}R\right) \nabla_{b)}\psi + 
{1\over 3}\, g_{ab}\,  \left\{ 2\, \sq^2 \psi + 6\,R^{cd} \,\nabla_c\nabla_d\psi
- 4\, R\, \sq \psi + (\nabla^c R)(\nabla_c\psi)\right\}\,.
\label{Fab}
\end{eqnarray}
Each of these two tensors are individually conserved and they have the local traces,
\begin{subequations}
\begin{eqnarray}
E_a^{\ a} &=& 2 \Delta_4 \varphi = E - \frac{2}{3} \sq R\,,\\
F_a^{\ a} &=& 2 \Delta_4 \psi = F = C_{abcd}C^{abcd}\,,
\label{Ftr}
\end{eqnarray}
\label{EFtraces}
\end{subequations} 
\vskip -.7cm
\noindent corresponding to the two terms respectively in the trace anomaly 
in four dimensions (with $\beta_i=0$). Results for the $E_{ab}$ tensor
in terms of the decomposition (\ref{metconf}), and the $F_{ab}$ tensor
in a slightly different notation were presented previously in refs.
\cite{AMM} and \cite{MazMot}, and partial results for $E_{ab}$ appear
also in \cite{BFS}.

It is straightforward to generalize the above development to include the
effects of any additional Weyl invariant terms in the trace anomaly,
transforming under (\ref{metconf}) by
\begin{equation}
\sqrt{-g}\,H_i = \sqrt{\bar g}\,\bar H_i\,.
\end{equation}
By making the replacement $F \rightarrow F + \sum_i\beta_i H_i/b$
in (\ref{auxvarphi}), (\ref{locaux}), and (\ref{SEF}), we observe that
$C[g;\varphi]$ should undergo the replacement,
\begin{equation}
C[g;\varphi] \rightarrow C[g;\varphi] + \sum_i \frac{\beta_i}{2b}  
\int\,d^4x\,\sqrt{-g}\ H_i \varphi\,,
\end{equation}
with the corresponding additional terms in $C_{ab}$ and $F_{ab}$ in
(\ref{Cten}), (\ref{ABCF}), (\ref{Fab}), and (\ref{Ftr}) to take
account of any additional contributions to the trace anomaly. Hence
no additional scalar auxiliary fields beyond $\varphi$ and $\psi$
are required in the general case. 

As in two dimensions, $S_{anom}$ in four dimensions contains massless 
scalar degrees of freedom which remain relevant at low energies and 
macroscopic distances, and must be retained in addition to the usual 
EFT local derivative expansion in order to take account of the 
non-decoupling of massless fields. In contrast, the Weyl invariant terms 
in (\ref{R2}) or $S_{inv}$ do not contain additional degrees of freedom 
of the gravitational field which are relevant at energies far below the 
Planck scale $M_{Pl}$, and are either marginally or strictly irrelevant 
in the infrared. 

In a given fixed spacetime background the solutions of the fourth order
differential equations (\ref{auxeom}) for the scalar auxiliary fields
$\varphi$ and $\psi$ may be found, and the results substituted into
the stress tensors $E_{ab}$ and $F_{ab}$. The freedom to add 
homogeneous solutions of (\ref{auxeom}) to any given inhomogeneous
solution corresponds to the freedom to change the Weyl invariant
part of the effective action and corresponding traceless parts
of the stress tensors, without altering their local traces 
(\ref{EFtraces}). The traceless terms in the stress tensor thus
depend on the state of the underlying quantum field(s) which 
can be fixed by specifying boundary conditions on the auxiliary
fields. When the spacetime is conformally flat or approximately so, 
the Weyl invariant action $S_{inv}$ can be taken to vanish, as it
does in the conformally related flat spacetime, and the low
energy effective action (\ref{Seff}) should become a good approximation
to the exact effective action (\ref{Sexact}). In order to test
this hypothesis we evaluate the corresponding anomalous stress tensor
(\ref{Tanom}) in conformally flat spacetimes for a number of different
macroscopic states.

\section{Stress-Energy Tensor in Conformally Flat Spacetimes}

\subsection{Flat Spacetime}

In flat Minkowski spacetime, $E = F = \sq R = 0$, and the auxiliary fields 
satisfy the homogeneous wave eqs., 
\begin{equation}
\Delta_4\Big\vert_{flat} \varphi = \sq^2 \varphi = 0 = \sq^2\psi\,.
\label{flateq}
\end{equation}
For the trivial solutions $\varphi = \psi = 0$ (or an arbitrary spacetime 
constant), the stress tensor vanishes identically. This corresponds to the 
ordinary Lorentz invariant vacuum state, with vanishing renormalized 
$\langle T_a^{\ b}\rangle$. Hence flat spacetime will continue
to satisfy the Einstein equations modified by the trace anomaly terms
(with vanishing cosmological term).

In flat spacetime the wave Eq. (\ref{flateq}) has plane wave solutions obeying 
the massless dispersion relation, $\omega_{\bf k}= \pm \vert{\bf k}\vert$. Since 
the wave operator here, $\sq^2$ is fourth order, there are two sets of massless 
scalar modes for {\it each} auxiliary field, $\varphi$ and $\psi$. These are new 
{\it local} scalar degrees of freedom for the gravitational field not present 
in classical general relativity, and give rise to scalar gravitational waves. 
These scalar gravitational waves of the augmented effective theory have as their 
sources the fourth order curvature invariants, $E, F$ and $\sq R$ which are 
quite small for most astrophysical sources, excepting perhaps those of 
cosmological origin. Hence in the flat space limit the scalar degrees of
freedom decouple completely. Even in the presence of matter, they couple only
indirectly through the effects of spacetime curvature, and only very weakly
at that. Localized sources act as only very weak generators of scalar 
gravitational radiation in the $\varphi$ and $\psi$ fields, and primordial 
scalar gravitational radiation, even if present, would be difficult to detect, 
except indirectly through the gravitational effects of the stress-energy (\ref{Tanom}).
The effects on the production of scalar gravitational waves from the coupling 
of $\psi$ to gauge fields through the $H_i$ terms in the trace (\ref{tranom}) 
is also possible, and merits a separate investigation.

For non-vanishing $\varphi$ the tensor $E_{ab}$ in flat space becomes
\begin{eqnarray}
&&E_{ab}\Big\vert_{flat} = -2\, (\nabla_{(a}\varphi) 
(\nabla_{b)} \sq \varphi) + 2 (\sq \varphi)(\nabla_a\nabla_b \varphi) 
+ \frac{2}{3}(\nabla_c \varphi)(\nabla^c\nabla_a\nabla_b\varphi)
- \frac{4}{3}\, (\nabla_a\nabla_c\varphi)(\nabla_b \nabla^c \varphi)
\nonumber\\
&& \qquad\qquad\qquad + {1\over 6}\, g_{ab}\, \left\{-3\, (\sq\varphi)^2 
+ \sq (\nabla_c\varphi\nabla^c\varphi) \right\}
- {2\over 3}\, \nabla_a\nabla_b \sq \varphi \,,
\label{Eflat}
\end{eqnarray}
in general curvilinear coordinates. We note that although the energy density 
for the scalar auxiliary fields is not positive definite, this does not lead 
to any instability at macroscopic distances in flat space, because of the
decoupling of the auxiliary fields in the flat space limit. In ref. \cite{AMM}
the physical state space of the conformal scalar fluctuations of $S_{anom}$ 
(proportional to $\varphi$), once decoupled from metric fluctuations, was shown 
to have positive norm, and be free of any tachyon or ghost modes. When
$S_{anom}$ is considered together with the classical Einstein-Hilbert terms,
because of the higher derivatives in the auxiliary field stress tensor, instability 
in flat space appears only at ultrashort Planck scale wavelengths \cite{AMolMot}, 
{\it i.e.} at the order of the cutoff scale and outside the range of validity of 
the low energy EFT approach. At the Planck scale the local expansion of the 
effective action in powers of $E^2/M_{Pl}^2$ clearly breaks down. For $E \ll M_{pl}$ 
where the low energy effective action (\ref{Seff}) can be applied, the auxiliary 
fields are coupled very weakly and indirectly to matter and propagate as very 
nearly non-interacting free field scalar excitations. This shows that empty flat 
space is stable to the effects of $S_{anom}$ on macroscopic distance scales,
and more subtle signatures of the presence of $S_{anom}$ in nearly flat spacetimes 
must be sought.

As a simple example of how the stress tensor (\ref{Eflat}) may be used to 
evaluate macroscopic semi-classical effects of quantum fields in flat space, 
consider the special solution to (\ref{flateq}),
\begin{equation}
\varphi = \frac{c_1}{2}\frac{z^2}{a^2}
\label{Cas}
\end{equation}
for some constant $a$ with dimensions of length and $c_1$ dimensionless. 
With a similar ansatz for $\psi = c_2 z^2/2a^2$ the total stress tensor 
of the anomaly (\ref{Tanom}) becomes
\begin{equation}
T_{ab}^{(anom)} = \frac{C}{3a^4}\ {\rm diag}\, ( -1, 1, 1, -3)\,,
\end{equation}
with $C= c_1^2 b' + 2bc_2^2$. 
This is the form of the stress tensor of the Casimir effect in region between 
two infinite parallel plates a distance $a$ apart in the $z$ direction. The
constants $c_1, c_2$ and $C$ depend on the boundary conditions which 
the quantum field(s) obey on the plates, and parameterize the discontinuities
in the components of the stress tensor there, but the generic form of the stress 
tensor appropriate for the geometry of two parallel conducting plates follows 
from the simple ansatz (\ref{Cas}) for the classical auxiliary fields
together with (\ref{Tanom}). Thus the low energy effective action of the trace 
anomaly in {\it curved} space correctly determines the form of the {\it traceless} 
components of the macroscopic quantum stress tensor for the parallel plate 
geometry in {\it flat} space.

If instead of (\ref{Cas}) we choose 
\begin{equation}
\varphi = \frac{c_1}{2}\,T^2 t^2
\label{Temp}
\end{equation}
with $\psi = c_2 T^2t^2/2$ the stress tensor (\ref{Tanom}) becomes
\begin{equation}
T_{ab}^{(anom)} = C\ \frac{T^4}{6}\ {\rm diag}\, (-3, 1, 1, 1)\,,
\end{equation}
with $C= -c_1^2 b' + 2bc_2^2$ which is the form of the stress tensor for 
massless radiation at temperature $T$, which is again traceless in flat space.
The same form for the anomalous stress tensor is also obtained by considering 
the general static spherically symmetric ansatz, $\varphi=\varphi(r)$. Then
\begin{equation}
\varphi = \frac{c_{-1}}{r} + c_0 + c_1 r + c_2 r^2\,,
\end{equation}
with $c_{-1} = 0$ to ensure a regular stress tensor at the origin.
We note that although a thermal state is a mixed state for the radiation field, 
it is a macroscopic pure coherent state for the gravitational auxiliary potentials 
at the semi-classical level, where fluctuations around the mean stress tensor 
$\langle T_{ab}\rangle = T_{ab}^{(anom)}$ are neglected.

For the four dimensional Rindler metric,
\begin{equation}
ds^2 = - \rho^2 d\eta^2 + d\rho^2 + dy^2 + dz^2
\end{equation}
and $\varphi = \varphi(\rho)$, $\Delta_4 \varphi = \sq^2 \varphi = 0$ has
the non-trivial solutions $\rho^2 \ln \rho, \rho^2$ and $\ln\rho$.
The first two solutions lead to stress tensors which are non-vanishing at 
infinity similar to the thermal state, while the $\ln \rho$ solution for $\varphi$
and $\psi$ gives
\begin{equation}
T_a^{\ b} = \frac{C}{\rho^4}\,{\rm diag} (-3, 1, 1, 1)\,.
\label{Rindf}
\end{equation}
Adding a term linear in $\eta$ to the solution for $\varphi(\rho)$ gives a 
stress tensor of the same form as (\ref{Rindf}) with a different constant $C$. 
In either case the form of the stress tensor (\ref{Rindf}) is that of the 
Fulling-Rindler state in four dimensions \cite{CanDeu}. Because of (\ref{delsingtwo}), 
the solution gives rise to a singularity in $\sq \varphi$ at $\rho =0$ 
proportional to $\delta(\rho)/\rho$, and therefore is a solution only in 
the distributional sense, as in the two dimensional case. The higher order 
derivatives in the Noether current in four dimensions produce higher order 
singularities at $\rho =0$, and the conformally transformed metric 
$e^{-\varphi} ds^2$ is singular at $\rho =0$ for any $c_1 \neq 0$.

\subsection{de Sitter Spacetime}

In conformally flat spacetimes with $g_{ab} = e^{2\sigma}\eta_{ab}$, one can
choose $\varphi = 2 \sigma$ and $\psi = 0$ to obtain the stress tensor
of the state conformally transformed from the Minkowski vacuum. In this state
$F_{ab}$ vanishes, and $\varphi$ can be eliminated completely
in terms of the Ricci tensor with the result \cite{BroCas,AEHMM,MazMot},
\begin{eqnarray}
E_{ab} &=& - 2\ ^{(3)}\hskip-.1cm H_{ab} - \frac{1}{9}\, ^{(1)}\hskip-.1cm H_{ab}
\nonumber \\
&=& \frac{2}{9} \nabla_a\nabla_b R + 2 R_a^{\ c}R_{bc} - \frac{14}{9}R R_{ab}
+ g_{ab} \left(-\frac{2}{9}\sq R - R_{cd}R^{cd} + \frac{5}{9} R^2\right)\,.
\label{Ecflat}
\end{eqnarray}
Thus all non-local dependence on boundary conditions of the auxiliary fields
$\varphi$ and $\psi$ vanishes in conformally flat spacetimes for the
state conformally mapped from the Minkowski vacuum. In the special case
of maximally $O(4,1)$ symmetric de Sitter spacetime, $R_{ab} = 3H^2 g_{ab}$
with $R = 12H^2$ a constant, the tensor $^{(1)}H_{ab}$ vanishes identically and
\begin{equation}
E_{ab}\Big\vert_{dS} = - 2\ ^{(3)}\hskip-.1cm H_{ab}\Big\vert_{dS} = 6 H^4 g_{ab}\,.
\end{equation}
Hence we obtain immediately the expectation value of the stress tensor
of a massless conformal field of any spin in the Bunch-Davies state 
in de Sitter spacetime,
\begin{equation}
T_{ab}\Big\vert_{BD,dS} = 6b' H^4 g_{ab} = -\frac{H^4}{960\pi^2}\,g_{ab}\,
(N_s + 11 N_f + 62 N_v)\,,
\label{BDdS}
\end{equation}
which is determined completely by the trace anomaly.

It is also possible to consider states which are not maximally $O(4,1)$
symmetric. For example, if de Sitter spacetime is expressed in the
spatially flat coordinates,
\begin{equation}
ds^2\Big\vert_{dS} = -d\tau^2 + e^{2H\tau}\, d\vec x^2\,
\label{flatdS}
\end{equation}
and $\varphi = \varphi(\tau)$, we obtain from (\ref{Tanom}) the stress-energy in
spatially homogeneous, isotropic states. Since in this maximally symmetric 
spacetime the $\Delta_4$ operator factorizes,
\begin{equation}
\Delta_4\Big\vert_{dS} \varphi = \left(\sq - 2H^2\right) (\sq \varphi) = 12H^4\,.
\label{deseom}
\end{equation}
it is straightforward to show that the general solution to this
equation with $\varphi = \varphi(\tau)$ is
\begin{equation}
\varphi(\tau) = 2 H \tau + c_0 + c_{-1} e^{-H\tau} + c_{-2} e^{-2H\tau} 
+ c_{-3} e^{-3H\tau} \,.
\label{destau}
\end{equation}
The first (inhomogeneous) term can be understood geometrically
from the fact that the conformal transformation,
\begin{equation}
e^{-\varphi_{_{BD}}}ds^2\Big\vert_{dS}= e^{-2H\tau}ds^2\Big\vert_{dS} 
= -d \eta^2 + d\vec x^2\,
\end{equation}
brings de Sitter spacetime to flat spacetime, with $\eta = -H^{-1} e^{-H\tau}$
the conformal time. This value of $\varphi = \varphi_{BD}(\tau) \equiv 2H\tau$
is exactly the one that gives the Bunch-Davies stress-energy (\ref{BDdS})
when substituted into (\ref{Eab}). When the full solution for $\varphi(\tau)$
of (\ref{destau}) is substituted into (\ref{Eab}) we obtain additional terms
in the stress tensor which are not de Sitter invariant, but which fall off
at large $\tau$, as $e^{-4H\tau}$. The stress tensor of this time behavior is 
traceless and corresponds to the redshift of massless modes with the equation 
of state, $p= \rho/3$. This is a special case of the detailed analysis of the 
stress tensor expectation value of a field of arbitrary mass in homogeneous, 
isotropic states in de Sitter spacetime of ref. \cite{AEHMM}. It is straightforward 
to generalize these considerations to spatially homogeneous and isotropic states 
in arbitrary conformally flat cosmological spacetimes with Robertson-Walker 
scale factor $a(\eta)$. Because of the reduction of the auxiliary field
stress tensor $E_{ab}$ to the local geometric form (\ref{Ecflat}) in
conformally flat spacetimes, it can have no significant effects in
states with Robertson-Walker symmetries when the curvature is much
less than the Planck scale.

States of lower symmetry in de Sitter spacetime may be considered 
by taking static coordinates,
\begin{equation}
ds^2 = -f(r) dt^2 + \frac{dr^2}{f(r)} + r^2 d\Omega^2\,,
\label{Sfour}
\end{equation} 
where 
\begin{equation}
f(r)\Big\vert_{dS} = 1 - H^2 r^2\,.
\end{equation} 
If we insert the ansatz $\varphi = \varphi(r)$ in (\ref{deseom}), the general 
$O(3)$ spherically symmetric solution regular at the origin is easily found:
\begin{equation}
\varphi(r)\Big\vert_{dS} = \ln\left(1-H^2r^2\right) + c_0 + \frac{q}{2} 
\ln\left({1-Hr\over 1+Hr}\right)
+ \frac{2c_{_H} - 2 - q}{2Hr} \ln\left({1-Hr\over 1+Hr}\right)\,.
\label{genphids}
\end{equation}
A possible homogeneous solution proportional to $1/r$ has been 
discarded, since it is singular at the origin.
An arbitrary linear time dependence $2Hpt$ could also be added to $\varphi (r)$,
{\it i.e.} $\varphi(r) \rightarrow \varphi(r,t) = \varphi(r) + 2Hpt$. 
The particular solution,
\begin{equation}
\varphi_{_{BD}}(r,t) = \ln\left(1-H^2r^2\right) + 2Ht = 2H\tau\,.
\label{phiBD}
\end{equation}
is simply the previous solution for the Bunch-Davies state we found
in the homogeneous flat coordinates (\ref{flatdS}). The priveledged
role of this fully $O(4,1)$ invariant state may be understood in
the static coordinates by recognizing that the line element (\ref{Sfour})
may be put into the alternate forms,
\begin{eqnarray}
ds^2 &=& f \left[-dt^2 + dr^{*\,2} + \frac{r^2}{f}\,d\Omega^2
\right]\,,\nonumber\\
&=& f \left[-dt^2 + \ell^2 \left(d\chi^2  + 
\sinh^2\chi\,d\Omega^2\right) \right]\,,\nonumber\\
&=& \frac{\ell^2 \,f}{\rho^2} \left[-d\rho^2 + \rho^2\left(d\chi^2  + 
\sinh^2\chi\ d\Omega^2\right) \right]\,,\nonumber\\
&=& \frac{f}{\rho^2} \left(-d\eta^2 + d\vec x^2\right)\,,\nonumber\\
&=&e^{\varphi_{BD}} (ds^2)_{flat}\,.
\label{descflat}
\end{eqnarray}
by means of the successive changes of variables, $\ell = H^{-1}$,
\begin{eqnarray}
\chi &=& Hr^* = H\int_0^r\,\frac{dr}{f(r)} = \frac{1}{2}\ln
\left(\frac{1+Hr}{1-Hr}\right)\,,\nonumber\\
t&=& - \ell \,\ln \rho\,,\nonumber\\
\eta &=& \ell \rho\, \cosh\chi\,,\nonumber\\
\vec x &=& \ell \rho\, \sinh\chi\ (\sin\theta \cos\phi\,, \sin\theta \sin\phi\,,
\cos\theta)\,.
\label{varch}
\end{eqnarray}
Thus de Sitter spacetime in static coordinates is conformally
related to flat spacetime by the conformal transformation $e^{\varphi_{BD}}$
with $\varphi_{BD}$ given by (\ref{phiBD}) as before.
The benefit of carrying this out explicitly in the static coordinates
is that the steps in (\ref{descflat}) can be repeated essentially
unchanged in the vicinity of {\it any} static Killing horizon, without
reliance on the special higher symmetries of de Sitter spacetime. Then
\begin{equation}
\varphi_h(r,t) = \ln f(r) \pm \frac{2}{\ell}\,t
\label{spphi}
\end{equation}
is the value of the auxiliary field which conformally transforms
the local neighborhood of the horizon to flat spacetime. In
de Sitter space this leading order conformal transformation
mapping the vicinity of the horizon to flat space is exact,
{\it cf.} (4.21).

In the general $O(3)$ symmetric state in de Sitter spacetime
the radial component of the Noether current,
\begin{eqnarray}
J^r &=& \nabla^r\sq\varphi -2H^2 \nabla^r \varphi - 
\frac{1}{2} \Omega^r\nonumber\\
&=& \frac{2Hq}{r^2}\,,
\end{eqnarray}
in the frame in which $\Omega^r = 8H^4r$. Thus the parameter $q$ 
coincides with the definition (\ref{charge}) and if non-zero gives 
rise to a delta function singularity for $\Delta_4 \varphi$
at the origin $r=0$. It is also the strength of the puncture
configurations (also called ``spikes") found in \cite{spike}.

In the general $O(3)$ symmetric state centered about the origin of
the coordinates $r=0$, the stress tensor is generally dependent on
$r$. In fact, it generally diverges as the observer horizon
$r= H^{-1}$ is approached. From (\ref{genphids}) we observe that
\begin{equation}
\varphi(r)\Big\vert_{dS} \rightarrow c_{_H} \ln\left(\frac{1-Hr}{2}\right)
 + {\cal O}(1-Hr) \qquad {\rm as} \qquad Hr \rightarrow 1\,,
\end{equation}
so that the integration constant $c_{_H}$ controls the singular behavior
at the observer horizon $r=H^{-1}$. As in the Schwarzschild case a
non-zero $c_{_H}$ produces a higher order singular distribution for
$\Delta_4 \varphi$ at the horizon. 

The second auxiliary field $\psi$ satisfies the homogeneous Eq., 
$\Delta_4 \psi = 0$, which has the general spherically symmetric
solution linear in $t$,
\begin{equation}
\psi(r,t)\Big\vert_{dS} = d_0 + 2Hp't + \frac{q'}{2} 
\ln\left({1-Hr\over 1+Hr}\right)
+ \frac{2d_{_H} - q'}{2Hr} \ln\left({1-Hr\over 1+Hr}\right)\,.
\label{genpsids}
\end{equation}
Note that the constant $d_{_H}$ enters this expression differently than
$c_{_H}$ enters the corresponding Eq. (\ref{genphids}), due to the
inhomogeneous term in (\ref{deseom}), which is absent from the $\psi$
equation. Since the anomalous stress tensor is independent of $c_0$ and $d_0$,
it depends on the six parameters $(c_{_H}, d_{_H}, q, q', p , p')$
in the general stationary $O(3)$ invariant state. However, in order to have 
no singularity of the stress tensor at the origin we must choose $q=q'=0$.
With no sources or sinks at the origin, the choice $q=q'=0$ then satisfies 
the zero flux condition,
\begin{equation}
T^{\ r}_t = -4H^2\,\left( bpq' + bp'q + b'pq\right) = 0
\label{dSflux}
\end{equation}
automatically, for any $p$ and $p'$. For the other components of the
stress tensor (\ref{Tanom}), the conditions of finiteness on the 
observer horizon at $r= H^{-1}$ are
\begin{eqnarray}
c_{_H} = 1\,, \qquad p = \pm 1\,, \qquad d_{_H} = \pm p'
\label{findes}
\end{eqnarray} 
which are satisfied by the Bunch-Davies state with $p=1$ and 
$d_{_H} = p'=0$. On the other hand, the local vacuum state at the 
origin ({\it i.e.} the analog of the Boulware vacuum \cite{Boul}
in de Sitter space) we should take $p=p'=0$. Since this is
inconsistent with (\ref{findes}), this state has a diverging 
stress-energy on the observer horizon. There is no analog of the 
Unruh state \cite{Unr} in de Sitter spacetime, since by continuity 
a flux through the future or past observer horizon at $r=H^{-1}$ would 
require a source or sink of flux at the origin $r=0$, a possibility we 
have excluded by (\ref{dSflux}) above.

\section{Near Horizon Conformal Symmetry and Stress Tensor}

The fact that the trace anomaly can account for even the tracefree parts 
of the stress tensor exactly in two dimensions and conformally flat
spacetimes in higher dimensions is due to the fact that $S_{inv}$
may be neglected in these cases. In these circumstances the low energy
effective action (\ref{Seff}) becomes a good approximation to the
exact effective action (\ref{Sexact}). However, as has been noted
by several authors \cite{CFT}, spacetimes with a static Killing horizon 
are locally conformally related to flat spacetime in the vicinity of the 
horizon. Because of this near horizon conformal symmetry we expect
the anomalous action (\ref{nonl}) to generate a stress tensor
(\ref{Tanom}) which is a good approximation to the quantum stress
tensor near an arbitrary Killing horizon.

To exhibit the conformal geometry of a static Killing horizon
consider the general static spherically symmetric spacetime 
of the form, (\ref{Sfour}) where the function $f(r)$ possesses a simple
zero ar $r=r_+$, {\it i.e.},
\begin{equation}
f(r) = \pm\frac{2}{\ell} (r-r_+) + {\cal O} (r-r_+)^2\,.
\label{fexp}
\end{equation}
Defining $r^*$ in the usual way, 
\begin{equation}
r^* = \int^r\, \frac{dr}{f(r)} \simeq \pm \frac{\ell}{2} \ln f + \dots\,,
\end{equation}
where the ellipsis denotes regular terms as $f \rightarrow 0$, and
\begin{equation}
\chi = - \frac{1}{2} \ln\left(\frac{\ell^2 f}{4r_+^2}\right) = 
\mp \frac{r^*}{\ell} +\dots\,,
\end{equation}
we find that near $r=r_+$,
\begin{equation}
\frac{r^2}{f(r)} \simeq \frac{\ell^2}{4} e^{2\chi} \simeq 
\ell^2 \sinh^2\chi\,.
\label{rovf}
\end{equation}
Hence we may repeat the steps (\ref{varch}) of the previous de Sitter
case for an arbitrary spherically symmetric static Killing horizon,
in the vicinity of the horizon. The length scale $\ell$ is also 
related to the surface gravity of the horizon by $\ell^{-1} = \kappa$.
Thus we have the conformal mapping (\ref{descflat}) of the near horizon 
metric (\ref{Sfour}) to the flat metric with $\varphi \simeq \varphi_h(r,t)$ 
given by (\ref{spphi}). Since this conformal transformation is singular at 
$r=r_+$, the exact effective action (\ref{Sexact}) is dominated by its 
anomalous contribution $S_{anom}$, and the stress-energy is dominated by
(\ref{Tanom}) in any state for which $\varphi$ and the corresponding
$T_a^{\ b}$ diverges on the horizon. Hence the near horizon conformal
symmetry (\ref{descflat}) may be used to calculate the exact behavior
of the divergent terms of the quantum stress-energy tensor from
(\ref{Tanom}) in any spherically symmetric static spacetime with
a Killing horizon. We illustrate this for the most important case
of Schwarzschild spacetime.

\subsection{Schwarzschild Spacetime}

In the four dimensional Schwarzschild geometry, $f(r) = 1 - {2M\over r}
\simeq r/2M - 1 + {\cal O}(r-2M)^2$. Hence $r_+ = 2M$ and $\ell = 4M$.
The conformal transformation that maps the near horizon geometry to
flat spacetime is therefore
\begin{equation}
\varphi_h(r,t) = \ln f \pm \frac{t}{2M}\,.
\label{conSch}
\end{equation}
Since Schwarzschild spacetime is not globally conformally flat
the conformal transformation of the near horizon geometry 
will receive subleading corrections to (\ref{conSch}).
To find these subleading terms we require the general solution to 
(\ref{auxeom}). Noting that Schwarzschild spacetime is Ricci flat,
we have
\begin{equation}
F\big\vert_{_S} = E\big\vert_{_S} = R_{abcd}R^{abcd}\Big\vert_{_S} = 
\nabla_a \Omega^a = {48\,M^2\over r^6}\,.
\label{Schtr}
\end{equation}
A particular solution of either of the inhomogeneous Eqs. (\ref{auxeom}) 
is given then by $\bar \varphi (r)$, with
\begin{equation}
{d\bar\varphi\over dr}\Big\vert_{_S} = -\frac{4M}{3r^2f} \ln \left(r\over 2M\right)
-\frac{1}{2M}\left(1 + \frac{4M}{r} \right) \,.
\end{equation}
Notice that this particular inhomogeneous solution to (\ref{auxeom}) is
regular as $r\rightarrow 2M$. The general solution of (\ref{auxeom}) for 
$\varphi =\varphi(r)$ away from the singular points $r= (0, 2M, \infty)$ is
easily found and may be expressed in the form \cite{BFS},
\begin{eqnarray}
{d\varphi\over dr}\Big\vert_{_S} &=&  {d\bar\varphi\over dr}\Big\vert_{_S} 
+ \frac{2M c_{_H}}{r^2f} + \frac{q-2}{4M^2r^2f} \int_{2M}^r\,dr\,r^2\,\ln f 
+ \frac{c_\infty}{2M}\, \left(\frac{r}{2M} + 1 + \frac{2M}{r}\right) \nonumber\\
&=&\frac{q-2}{6M}\,\left(\frac{r}{2M} + 1 + \frac{2M}{r}\right) 
\ln \left(1 - \frac{2M}{r}\right) - \frac{q}{6r}\left[\frac{4M}{r-2M} 
\ln \left(r\over 2M\right) + \frac{r}{2M} + 3 \right] \nonumber\\
&& \qquad  - \frac{1}{3M} - \frac{1}{r} + \frac{2M c_{_H}}{r (r-2M)} 
+ \frac{c_\infty}{2M}\, \left(\frac{r}{2M} + 1 + \frac{2M}{r}\right)
\label{phipS}
\end{eqnarray}
in terms of the three dimensionless constants of integration,
$c_{_H}$, $c_\infty$, and $q$. This expression has the limits,
\begin{subequations} 
\begin{eqnarray}
&&\hskip-1.3cm {d\varphi\over dr}\Big\vert_{_S} \rightarrow {c_{_H}\over r-2M}
+ {q-2\over 2M} \,\ln\left({r\over 2M} - 1\right) - {1\over 2M}
\left(3 c_\infty - c_{_H} - q - \frac{5}{3}\right) + \dots,\qquad r
\rightarrow 2M, 
\label{Schlima}\\
&&\hskip-1.3cm {d\varphi\over dr}\Big\vert_{_S} \rightarrow {c_\infty r \over 4M^2} + 
\frac{2c_\infty - q}{4M} + \frac{c_\infty}{r} - \frac{2M}{3r^2}\,q\, 
\ln \left(r\over 2M\right) + \frac{2M}{r^2}\left[c_{_H} - \frac{7}{18}
(q - 2)\right] + \dots,\,r \rightarrow \infty .
\label{Schlimb}
\end{eqnarray}
\end{subequations}

Hence $c_{_H}$ controls the leading behavior as $r$ approaches the 
horizon, while $c_\infty$ controls the leading behavior as $r\rightarrow \infty$, 
which is the same as in flat space. The leading behavior at the horizon is 
determined by the homogeneous solution to (\ref{auxeom}), $c_{_H} \ln f = c_{_H}
\ln (-K^aK_a)^{\frac{1}{2}}$ where $K = \partial_t$ is the timelike Killing
field of the Schwarzschild geometry for $r>2M$. To the general spherically
symmetric static solution (\ref{phipS}) we may add also a term linear
in $t$, {\it i.e.} we replace $\varphi(r)$ by
\begin{equation}
\varphi (r,t) = \varphi (r) + \frac{p} {2M}\,t\,.
\label{tSch}
\end{equation}
Linear time dependence in the auxiliary fields is the only allowed
time dependence that leads to a time-independent stress-energy.
We see that the conformal transformation to flat space near the horizon
(\ref{conSch}) corresponds to the particular choice $c_{_H} = \pm p = 1$,
leaving the subdominant terms in (\ref{Schlima}) parameterized
by $q$ and $c_\infty$ undetermined.

The topological charge $q$ may be identified from
\begin{equation}
\nabla^r \sq \varphi = - \frac{8M^2}{r^5} + \frac{q}{2Mr^2}\,,
\end{equation}
so that $q$ is given by (\ref{topcharge}) in the Euclidean signature Schwarzschild
geometry in the frame in which $\Omega^r = -16M^2/r^5$ and $\Omega^t = 0$.
A non-zero $q$ gives $\Delta_4 \varphi$ a delta function singularity at
the origin. Non-zero $c_{_H}$ or $c_\infty$ produce higher order singular
distributions in $\Delta_4 \varphi$ at the horizon or infinity respectively.
We note that unlike in flat space there are logarithmic terms in $\varphi(r)$ and 
no value of $q$ which can eliminate the subleading logarithmic behavior of
$\varphi'$ at both $r=2M$ and $r=\infty$. 

Since the equation for the second auxiliary field $\psi$ is identical to that 
for $\varphi$, its solution for $\psi = \psi(r,t)$ is of the same form as 
(\ref{phipS}) and (\ref{tSch}) with four new integration constants, 
$d_{_H}, d_\infty$, $q'$ and $p'$ replacing $c_{_H}, c_\infty$, $q$,
and $p$ in $\varphi(r,t)$. Adding terms with any higher powers of $t$ or 
more complicated $t$ dependence produces a time dependent stress-energy tensor. 
Inspection of the stress tensor terms in (\ref{Tanom}) also shows that it does not 
depend on either a constant $\varphi_0$ or $\psi_0$ but only the derivatives of 
both auxiliary fields in Ricci flat metrics such as Schwarzschild spacetime. For 
that reason we do not need an additional integration constant for either of the 
fourth order differential equations (\ref{auxeom}).

With the general spherically symmetric solution for $\varphi(r,t)$ and 
$\psi(r,t)$, we can proceed to compute the stress-energy tensor in a stationary,
spherically symmetric quantum state. For example the Boulware state may be 
characterized as that state which approaches the flat space vacuum as rapidly 
as possible as $r\rightarrow \infty$ \cite{Boul}. In the flat space 
limit this means that the allowed $r^2$ and $r$ behavior in the auxiliary
fields ($r$ and constant behavior in their first derivatives) must be set
to zero. Inspection of the asymptotic form (\ref{Schlimb}) shows that
this is achieved by requiring
\begin{subequations}
\begin{eqnarray}
c_{\infty} = d_{\infty} &=& 0\qquad {\rm and}\\
q=q' &=& 0\,.\qquad \qquad {\rm (Boulware)}
\label{Boulcond}
\end{eqnarray}
\end{subequations}

\noindent If we set $p=p'=0$ as well, in order to have a static
anasatz for the Boulware state, then the remaining two constants 
$c_{_H}$ and $d_{_H}$ are free parameters of the auxiliary fields,
and lead to a stress-energy which generically has the divergent behaviors,
\begin{equation}
s^{-2}, \quad s^{-1}, \quad \ln^2 s, \quad {\rm and} \quad \ln s,
\quad {\rm as} \quad s\equiv \frac{r-2M}{M} \rightarrow 0\,,
\label{singS}
\end{equation}
on the horizon. It is not possible to cancel all four of these
divergent behaviors with the two remaining free parameters.
As in the two dimensional case, it is not possible to have 
auxiliary fields and stress-energies which both fall off at
infinity and are regular on the horizon. This macroscopic
effect of the trace anomaly is a result of the non-trivial
global topology of the Schwarzschild metric, notwithstanding the
smallness of local curvature invariants at the horizon. 

The linear time dependence (\ref{tSch}) is consistent with the singular 
behaviors (\ref{singS}) of the stress-energy on the horizon. Since the 
stress-energy in the Boulware state diverges as $s^{-2}$ in any case, 
we can match this leading divergence in the anomalous stress tensor by 
adjusting $c_{_H}$ and $d_{_H}$ appropriately. One then has a one parameter 
fit to the numerical data of \cite{JenMcOtt}. The results of this fit of 
(\ref{Tanom}) with $c_{_H}$ and $d_{_H}$ treated as free paramters are 
illustrated in Figs. 1, for all three non-zero components of the stress
tensor expectation value of a massless, conformally coupled scalar field 
in the Boulware state. The best fit values plotted were obtained 
with $c_{_H} = -\frac{7}{20}$ and $d_{_H} = \frac{55}{84}$. 

For comparison purposes, we have plotted also the analytic approximation 
of Page, Brown, and Ottewill \cite{Pag82,BroOtt,BrOtP} (dashed curves in 
Figs. 1). We observe that the two parameter fit with the anomalous stress 
tensor in terms of the auxiliary $\varphi$ and $\psi$ fields is more 
accurate than the approximation of refs. \cite{Pag82,BroOtt,BrOtP} for 
the Boulware state. In the latter case the stress-tensor is approximated 
by making a special conformal transformation
$\sigma = \frac{1}{2}\ln f = \frac{1}{2}\ln (-K^aK_a)$ to the optical 
metric, obtained from the Schwarzschild line element (\ref{Sfour}) by 
dividing by $f(r) = 1 - \frac{2M}{r}$. This procedure was also motivated 
by the form of the trace anomaly, in that for ultrastatic metrics with 
the timelike Killing field $K^a$, the optical metric conformally related
it has vanishing trace anomaly \cite{BroOtt}. However, unlike the general 
form of the anomalous stress tensor in (\ref{Tanom}), which involves 
solving linear equations for two independent auxiliary fields, the 
approximation of refs. \cite{Pag82,BroOtt,BrOtP} is applicable only 
to special cases, such as Schwarzschild geometry which are both 
ultrastatic and Ricci flat, and admit a particular conformal 
transformation to the static optical metric. In contrast the approach 
based on the auxiliary field anomalous effective action does not require 
any special properties of the background spacetime, and because of the 
near horizon conformal symmetry gives the general form of the divergences 
of the stress-energy on horizon. An important practical difference between 
these approximations and that considered here is that they yield simple 
algebraic approximations to the stress-energy, whereas (\ref{Tanom}) 
generally contains logarithmic terms as well. Evidently the general form 
of the effective stress-energy tensor due to the anomaly in terms of two 
auxiliary fields and the freedom to add homogeneous solutions to the linear 
Eqs. (\ref{auxeom}) allows for a better approximation to the exact stress 
tensor in the Boulware state than that of refs.\cite{Pag82,BroOtt,BrOtP} 
near the Schwarzschild horizon.

The stress-energy diverges on the horizon in an entire family of states 
for generic values of the eight auxiliary field parameters 
$(c_{_H}, q, c_{\infty}, p; d_{_H}, q', d_{\infty}, p')$, in addition 
to the Boulware state. Hence in the general allowed parameter space of 
spherically symmetric macroscopic states, horizon divergences of the 
stress-energy are quite generic, and not resticted to the Boulware state. 
On the other hand, the condition (\ref{finT}) that the stress-energy om
the horizon be finite gives four conditions on these eight parameters, in 
order to cancel the four possible divergences listed in (\ref{singS}). 
Referring to the explicit form of the stress-energy components in static 
spherically symmetric spacetimes given in the Appendix, these four conditions 
can be satisfied in several different ways. The simplest possibility with 
the minimal number of conditions on the auxiliary field parameters is via
\begin{subequations}
\begin{eqnarray}
&& (2b + b') c_{_H}^2 + p (2b p' + b'p) = 0\,,\qquad (s^{-2})\,,\\
&& (b + b') c_{_H} = bd_{_H}\,, \qquad (s^{-1})\,,\\
&& q=q' = 2 \,,\qquad (\ln^2s \quad {\rm and} \quad \ln s)\,. 
\qquad{\rm (horizon\ finiteness)}
\end{eqnarray}
\label{fincond}
\end{subequations}
\vskip-1cm
\noindent It is noteworthy that this cancellation requires a
non-zero topological charge $q=2$, much as in the two dimensional
Schwarzschild case, and again unlike the Fulling-Rindler case
in either two or four dimensions. The $q=2$ condition may be
understood from the fact that it cancels the logarithmic
divergence of $\ln f$ in the second form of (\ref{phipS}),
leaving however a logarithmic singularity and non-trivial
topological charge at both the origin and infinity.

Also noteworthy is that the the auxiliary field configurations which 
lead to diverging stress tensors on the horizon have {\it finite action} 
(\ref{allanom}). This is clear from the fact that
\begin{equation}
\sq \varphi \rightarrow \frac{q-2}{4M^2}\, 
\ln\left(\frac{r}{2M} -1\right) + const.\qquad {\rm as} \qquad
r\rightarrow 2M\,.
\end{equation}
Hence both $(\sq \varphi)^2$ and $(\sq\varphi)(\sq\psi)$ are proportional 
to $\ln^2(r/2M - 1)$, which is integrable with respect to the measure 
$r^2 dr$ as $r\rightarrow 2M$, and the integrals in (\ref{SEF}) converge 
at the horizon for the general solution (\ref{tSch}) of the auxiliary field 
equations. This is not the case at infinity, unless the four constants 
$c_{\infty}, d_{\infty}, q$ and $q'$ are all required to vanish. 
From (\ref{phipS}) in this case the auxiliary fields fall off at large 
distances $r \gg 2M$ similarly to the scalar potential of Newtonian gravity, 
or Brans-Dicke theory \cite{BraDic}. Because of the very different way 
they couple to spacetime curvature, the auxiliary fields of $S_{anom}$ give 
only a very weak long-range interaction between massive bodies, that falls 
off very rapidly with distance (at least as fast as $r^{-7}$). 
Thus from consideration of the general form of the anomalous action and 
auxiliary fields in the vicinity of the Schwarzschild horizon and their 
falloff at infinity, we find no {\it a priori} justification for excluding 
generic states with stress-energy tensors that grow without bound as the 
horizon is approached. In such states the backreaction of the stress-energy 
on the geometry will be substantial in this region and lead to qualitatively 
new macroscopic effects on the horizon scale.

If instead of requiring rapid falloff at infinity we require 
regularity of the stress-energy on the horizon, {\it i.e.}
conditions (\ref{fincond}), which is a four parameter restriction 
of the eight parameters in the auxiliary fields of the form 
(\ref{phipS}) and (\ref{tSch}). These are the conditions 
appropriate to the Hartle-Hawking-Israel and Unruh states \cite{HarHaw,Unr}.
Because of the linear time dependence (\ref{tSch}) we also
obtain a no-zero flux $\langle T_t^{\ r}\rangle$ in general.
For the Hartle-Hawking state the vanishing of this flux gives 
a fifth condition, {\it viz.}
\begin{equation}
b(q p' + q' p) + b' pq = 0\,, \quad {\rm (zero\ flux)}
\label{flux}
\end{equation}
on the parameters of the auxiliary fields. Of the remaining three 
parameters two can be fit by the finite values of 
$\langle T_t^{\ t}\rangle$ on the horizon and at infinity
respectively. This leaves one final parameter free to adjust for a 
best fit of the form of all three components of the stress-energy 
tensor at intermediate points. 

Alternately, the conformal transformation (\ref{conSch}) could be used 
to fix the leading order behavior of the auxiliary field $\varphi$ near 
the horizon, {\it i.e.} $c_{_H} = 1 = \pm p$. Then (\ref{flux}) fixes $p'$ 
which with $q=q'$ becomes redundant with the first of the conditions 
(\ref{fincond}). Hence only $c_\infty$ and $d_\infty$ are left undetermined, 
and no free parameters at all are left if the values of $\langle T_t^{\ t}\rangle$ 
on the horizon and at infinity are fixed. Using the first method with one free 
parameter, the results for the $\langle T_t^{\ t}\rangle$ component of the 
scalar, Dirac spin $\frac{1}{2}$ and electromagnetic fields are shown in 
Figs. 2 (solid curves), for typical values of the parameters. The data 
points plotted in Figs. 2a and 2b are the numerical results of the direct 
quantum field theory calculations for spin $0$ and $\frac{1}{2}$ of 
refs. \cite{How} and \cite{CHOAG} respectively. In the spin $1$ case the 
dashed curve represents the numerical results of ref. \cite{JenOtt} in an 
accurate analytic polynomial fit to their data provided by the same authors. 

We note that the energy density in the Hartle-Hawking state computed from 
the general form of the anomaly is not in especially good agreement with 
the direct numerical calculations from the underlying quantum field theory. 
This is in accordance with the general discussion of the exact and low energy 
effective actions of the previous section. In states which have no divergences
on the horizon there is no particular reason to neglect the Weyl invariant 
terms $S_{inv}$ in the exact effective action, and the stress-energy tensor 
it produces would be expected to be comparable in magnitude to that from 
$S_{anom}$, and both contributions are negligibly small on macroscopic
scales. However in states such as the previous Boulware example, the diverging 
behavior of the stress tensor near the horizon is captured accurately by the 
terms in (\ref{Tanom}) arising from the anomaly, which have the same generically 
diverging behaviors as the quantum field theory expectation value 
$\langle T_a^{\ b}\rangle$. Even in states for which the expectation value
remains regular on the horizon, the auxiliary field anomalous stress-energy
can be matched accurately to the finite value of $\langle T_a^{\ b}\rangle$
on the horizon with suitable choice of integration constants in (\ref{phipS}).

For comparison plotted also in Figs. 2a and 2b is the analytic approximation 
to $\langle T_t^{\ t}\rangle$ of Frolov and Zel'nikov (FZ) \cite{FroZel}. This 
latter approximation gives an improvement over that of \cite{Pag82,BroOtt,BrOtP} 
since it permits adjustment of the value of the stress-energy on the horizon to 
the numerical data, rather than fixing it to an incorrect value. Once that is 
done, the FZ approximation automatically gives the correct asymptotic value
of the stress-energy in the Hartle-Hawking state at infinity, but contains 
no additional free parameters. For the spin $\frac{1}{2}$  case the FZ 
approximation is a significant improvement of that of PBO which disagrees 
with the numerics even in the sign of $\langle T_t^{\ t}\rangle$ on the horizon. 
Although it is a noticeably better fit to the numerical data than (\ref{Tanom}), 
the FZ approximation, like that of PBO, relies on the existence of a timelike 
Killing field and both are therefore highly specialized to certain spacetime 
backgrounds such as the Schwarzschild black hole. In contrast the stress-energy 
tensor (\ref{Tanom}) arising from $S_{anom}$ in the low energy EFT does not 
require the existence of a timelike Killing field, and can be computed 
in principle for arbitrary spacetimes.

For the time asymmetric Unruh state the zero flux condition (\ref{flux}) 
is replaced by the properly normalized finite flux condition,
\begin{equation} 
\langle T_t^{\ r}\rangle\Big\vert_{_U} = -\frac{L}{4\pi r^2}\,
\label{lumflux}
\end{equation}
where the luminosity, $L$ is given by
\begin{equation}
L = \frac{\pi}{M^2} \left(b q p' + b q' p + b' pq\right)
\end{equation}
in terms of the auxiliary field parameters. Otherwise one can proceed as 
in the previous Hawking-Hartle state by fixing two of the three remaining
auxiliary field parameters by the values of $\langle T_t^{\ t}\rangle$ 
on the future horizon and at infinity which are known numerically for
the scalar and electromagnetic fields in the Unruh state from the work
of refs. \cite{Pag76} and \cite{Elst,JenMcOttBR}. This leaves again 
one parameter free to fit. The results of that fit for the conformally 
coupled scalar and electromagnetic fields in the Unruh state are shown in 
Figs. 3 (solid lines). The dashed curves are accurate polynomial fits of 
\cite{Vis} and \cite{Matyj} to the numerical data of \cite{Elst} and 
\cite{JenMcOttBR} for the spin $0$ and $1$ fields respectively. There is 
at present no numerical calculation of the stress-energy tensor of the 
massless Dirac field in the Unruh state with which to compare. 

\subsection{Other Spacetimes with Horizons}

The foregoing analysis of the anomalous action and stress tensor in
the Schwarzschild background may be extended to other spacetimes with
metrics of the form (\ref{Sfour}). The electrically charged Reisner-Nordstrom 
black hole for which $f(r) = 1 - 2M/r + e^2/r^2$ is an example in which
Eqs. (\ref{fexp})-(\ref{rovf}) may also be applied. The extreme Reisner-Nordstrom
geometry in which $e=M$ is particularly interesting. In that case the two
zeroes of $f(r)$ at $r_{\pm}$ coincide and $f$ vanishes quadratically.
The possible singularities of the stress-energy on the extreme Reissner-Nordstrom
horizon are correspondingly more severe, behaving like $s^{-3}$. Nevertheless
a power series expansion near $s=0$ shows that it is possible to obtain a regular 
stress-energy on the horizon with (\ref{Tanom}) even in the extreme 
Reisner-Nordstrom case with appropriate choice of auxiliary field integration 
constants, in agreement with the direct numerical results of \cite{CHOAG}.
Thus the anomaly induced effective action can reproduce the
correct behaviors of the quantum stress tensor also in this case, in contrast
to the approximation of \cite{Pag82,BroOtt,BrOtP} which predicts a divergent 
stress-energy on the horizon of an extreme Reisner-Nordstrom black hole as the 
only possibility, As it is not possible to solve the auxiliary field equations 
analytically in the black hole geometries when $e\neq 0$, and a numerical
solution is required, we postpone a full discussion of these results to a 
future publication \cite{AMV}.

Finally, this study the anomalous effective action, auxiliary fields, and 
stress-energy could be extended to rotating Kerr black hole geometries. 
The previous analytical approximations of \cite{FroZel,Pag82,BroOtt,BrOtP}
do not apply to this case. However there is no problem of principle in
computing the stress tensor (\ref{Tanom}) in stationary spacetimes 
possessing only axial symmetry. Because of the lower symmetry, solutions
of the auxiliary field equations (\ref{auxeom}) which are functions of both
$r$ and $\theta$ can be sought. Hence the solutions characterizing
macroscopic stationary quantum states with axial symmetry are
considerably richer than the Schwarzschild case. However, a simple 
spherically symmetric homogeneous solution of the second order 
equation $\sq \varphi = 0$ which generalizes the $\ln f$ solution
found in the the Schwarzschild and de Sitter cases is
\begin{equation}
\varphi(r) \propto \ln \left(\frac{r-r_+}{r-r_-}\right)\,,
\label{Kerrlog}
\end{equation}
for
\begin{equation}
r_{\pm} = M \pm \sqrt{M^2 -a^2}\,,
\end{equation}
with $a$ the the angular momentum per unit mass of the Kerr solution.
Since (\ref{Kerrlog}) diverges logarithmically on the horizon at
$r=r_+$, similar conformal behavior of the stress tensor as 
$r\rightarrow r_+$ is expected as that found in the Schwarzschild case, 
which clearly merits a separate investigation.

\section{Conclusions}

The suggestion that general relativity is a low energy effective field 
theory, comparable to the chiral EFT of pions in the strong interactions 
has been expressed previously \cite{Dono}. In this paper we have extended 
the EFT framework from the weak field expansion around flat space
considered in \cite{Dono} to arbitrary curved spacetimes, provided
only that the curvature invariants are small compared to the Planck scale.
The EFT expansion in inverse powers of the high energy cutoff scale
of order $M_{Pl}$ requires a systematic classification of relevant and 
irrelevant operators in the general coordinate invariant effective action. 
According to this classification, based on the behavior under global Weyl 
rescalings \cite{MazMot}, the classical Einstein-Hilbert action should be 
supplemented with certain well-defined non-local terms associated with the 
trace anomaly of massless fields. These additional terms are non-local in 
terms of the metric, {\it cf.} (\ref{nonl}) but nevertheless because they 
scale logarithmically with distance are relevant in the infrared. The 
anomalous terms are unsuppressed by any inverse power of the 
ultraviolet cutoff scale $M_{Pl}$, and do not decouple for $E \ll M_{Pl}$.
Hence they are required in the Wilson effective action (\ref{Seff})
for gravitational physics at macroscopic distances $L \gg L_{Pl}$.
This constitutes an non-trivial infrared modification of Einstein's theory
which is completely consistent with the Equivalence Principle.

The local form of the anomaly introduces two scalar auxiliary fields,
$\varphi$ and $\psi$, into the EFT for low energy gravity, one for each 
of the non-trivial co-cycles of the Weyl group. These scalar fields
satisfy the fourth order massless wave equations (\ref{auxeom}), and
are only very weakly coupled to ordinary matter and radiation.
In the flat space limit they decouple entirely. The auxiliary fields 
are new local scalar degrees of freedom of the gravitational field, 
not contained in classical general relativity.

The anomalous terms in the effective action also generate two new
conserved tensors (\ref{Eab}) and (\ref{Fab}), local in the auxiliary
fields, which contribute to the stress-energy tensor of low energy
gravity (\ref{Tanom}), as a source for Einstein's equations in vacuo.
Since they arise from the long range effects of the fluctuations
of massless fields, and require boundary conditions for their
complete specification, these additional terms carry information 
about the global macroscopic effects of quantum matter. In the 
approximation where the auxiliary scalar fields are treated as 
classical potentials, they may be viewed as macroscopic order 
parameter fields of low energy gravity. 

Because of the conformal properties of spacetime horizons, the 
anomalous terms in the effective action and stress-energy become 
important in their vicinity, characterizing the allowed behaviors of 
the stress tensor there. Near an horizon the most important terms in the 
effective action (\ref{Seff}) are those which scale positively in the 
{\it ultraviolet} short distance limit under conformal rescalings, while 
nevertheless keeping local curvature invariants fixed and small compared to 
the Planck scale. Only the $S_{anom}$ term in (\ref{Seff}) has this property, 
scaling logarithmically in both the ultraviolet and infrared limits. Because 
the anomalous terms depend non-locally on the geometry through the two 
auxiliary scalar fields $\varphi$ and $\psi$, they can be significant 
near horizons even for very much sub-Planckian local curvatures, where the 
EFT description should continue to apply. Indeed although local curvature 
invariants remain small on horizons, the auxiliary scalar fields depend
in general on coordinate invariant global quantities such as $\ln (-K^aK_a)$ 
which diverge on the horizon when the spacetime possesses a timelike Killing
field $K^a$ which becomes null there. Whether or not this divergence 
actually occurs in the stress-energy of a given state is a matter of 
boundary conditions fixed globally over the entire spacetime. The 
existence of a topological current and associated Noether charge are
characteristic of these global effects of the quantum state. The generic 
singular behavior of the stress-energy on horizons is thereby characterized 
in terms of spacetime invariant scalar order parameters in the low energy EFT, 
which is completely consistent with general coordinate invariance.

In the (semi-)classical EFT framework where both the metric and auxiliary 
scalars are treated as classical fields, Eqs. (\ref{auxeom})
may be solved classically and provide a new method to calculate the
approximate quantum stress tensor of conformal matter by classical 
means, bypassing the standard but cumbersome procedure of field 
quantization and renormalization of the quantum stress tensor.
When states for which the stress-energy does not diverge on the horizon
are considered, approximations such as those in \cite{FroZel,Pag82,BroOtt,BrOtP},
developed specifically for static backgrounds with a timelike Killing
field yield a more accurate approximation to the expectation value of the 
stress-energy tensor. We conclude that in states with a finite stress-energy 
on the horizon the low energy Wilson effective action (\ref{Seff}) is not a 
particularly good approximation to the exact quantum effective action 
(\ref{Sexact}), as indeed there is no reason why it should be in such states
where quantum effects are negligibly small in any case.

On the other hand when the scalar potentials diverge on the horizon, the 
associated terms in the stress-energy yield a good approximation to the 
vacuum expectation value of the stress-energy tensor. Compared to previous 
analytic approximations (\ref{Tanom}) with (\ref{auxeom}) has the considerable 
advantage of being completely general, and giving all possible allowed singular 
behaviors of the stress tensor on horizons. In this case the auxiliary scalar 
potentials may be viewed as horizon order parameters. Even when the stress-energies 
diverge, they are characterized by auxiliary fields with finite action, suggesting 
that such states should not be excluded {\it a priori} from physical consideration.

We conclude that the effective action and stress-energy of the anomaly
provides a consistent infrared modification of general relativity that
incorporates long-range macroscopic quantum coherence and entanglement
effects in a natural way. This low energy modification is actually required
on general EFT grounds. The auxiliary fields of the anomalous action and
stress-energy may be viewed as macroscopic condensates.

The study of their effective stress-energy tensor in several special 
fixed spacetime backgrounds presented here clearly only begins a systematic 
treatment of their many possible other effects. In addition to extension to 
other backgrounds, particularly those with horizons, and a study of the 
scalar gravitational waves and long range effects these scalar potentials may 
generate, a rich variety of new solutions are to be expected when the $T_{ab}$ of 
Eqs. (\ref{Tanom}) is used as fully dynamical sources for Einstein's equations. 
Whether these effects can be used to test the theory against astrophysical
observations is a most important issue for further investigation.
In the case of divergent behavior of the stress-energy near the classical
horizon a significant deviation of the geometry from the vacuum solutions of 
the purely classical Einstein equations is predicted. This raises the distinct 
possibility of qualitatively changing the global structure of both classical black 
hole and cosmological solutions \cite{PNAS}. The existence of an abelian conserved 
Noether charge in the EFT suggests comparison with a number current similar to 
that of non-relativistic many-body systems at a microscopic level \cite{Maz}.

\vskip .5cm
\centerline{{\it Acknowledgments}}
\vskip .5cm

The authors gratefully acknowledge extensive discussions with P. R.
Anderson on several parts of this work, and would like to thank him
and Eric Carlson for the use of his numerical data from ref. \cite{CHOAG} 
in Fig. 2b. We also thank G. A. Newton for helping to develop the
MathTensor \cite{matht} code which we used to check Eqns. (\ref{Eab}) 
and (\ref{Fab}) and generate those in the Appendix. We thank P. O. Mazur 
for critical reading of the manuscript and several helpful suggestions 
prior to its submission. R. V. gratefully acknowledges support while at 
Los Alamos from the Univ. of Calif. Insitute for Nuclear and Particle 
Astrophysics and Cosmology.


\appendix
\section{Stress-Energy in Static, Spherical, Vacuum Spacetimes}

We give in this Appendix the explicit forms for the diagonal components
of the two conserved tensors $E_a^{\ b}$ and $F_a^{\ b}$, {\it cf.} Eqns.
(\ref{Eab}) and (\ref{Fab}), for static, spherically symmetric spacetimes
with line elements of the form (\ref{Sfour}). The auxiliary fields are
assumed to be of the form,
\begin{subequations} 
\begin{eqnarray}
\varphi &=& \varphi (r) + 2\,p\,\kappa\, t\,,\\
\psi &=& \psi (r) + 2\,p'\,\kappa\, t\,,
\end{eqnarray}
\end{subequations}
with $\kappa$ the surface gravity at the horizon where $f$ vanishes.
All components of the two tensors were generated by MathTensor programs \cite{matht},
which were separately checked for satisfying the conservation and 
trace conditions (\ref{EFtraces}).

The time components of the two tensors so obtained are:
\begin{eqnarray}
&&\hskip -.6cm E^{\ t}_t =  \frac{{f}^2}{6}\left(
2\,\varphi'''\varphi'
- {\varphi''}^2\right)
+ \frac{4f}{3}\left(f' 
- \frac{f}{r}\right)\varphi''\varphi'
+ \frac{1}{3} \left(2ff'' 
+ {f'}^2  
- \frac{2f(1+2f)}{r^2}\right){\varphi'}^2
+ \frac{2p^2\kappa^2 {f'}^2}{f^2}\nonumber\\
&& 
- \, \frac{8p^2\kappa^2}{3f}\left(f'' 
+ \frac{f'}{r} + \frac{1-f}{r^2}\right)
- \frac{ff'}{3}\, \varphi'''
- \frac{2}{9} \left[\frac{ff''}{2} 
+ 3 {f'}^2 
+ \frac{8ff'}{r}
+ \frac{14f(1-f)}{r^2}\right] \varphi''\nonumber \\
&&
\hskip 1.2cm - \frac{2}{9} \left[\frac{ff'''}{2}
+ 2f'f''   
+ \frac{9ff''+8{f'}^2}{r} 
+ \frac{(5 - 17f)f'}{r^2} 
+ \frac{6f(1-f)}{r^3}\right] \varphi'\nonumber\\
&&
\hskip -.1cm + \,\frac{4}{9}\left[\frac{ff'''' + f'f'''}{2}
+ \frac{2f'f'' + 3ff'''}{r}  
+ \frac{f''(4f-3) + 2{f'}^2}{r^2}
+ \frac{2f'(1 - 2f)}{r^3}
- \frac{2f(1-f)}{r^4} \right]\,,
\end{eqnarray}
and
\begin{eqnarray}
&&\hskip-1cm F^{\ t}_t =
\frac{f^2}{3}(\varphi'''\psi' + \varphi'\psi''' - \varphi''\psi'')
+ \frac{4f}{3} \left(f' - \frac{f}{r}\right)\, 
(\varphi''\psi' + \varphi'\psi'')
+ \frac{2}{3}\left[2ff'' 
+ {f'}^2 
- \frac{2f(1+2f)}{r^2} \right]\varphi'\psi' \nonumber \\
&&
\hskip-.4cm + \,\frac{ 4 pp'{\kappa}^2{f'}^2}{{f}^2} 
- \frac{16 pp'{\kappa}^2}{3f}  \left(f''
+\frac{f'}{r} + \frac{1-f}{r^2}\right)
- \frac{ff'}{3}\psi'''
-\frac{2}{9}\left[\frac{ff''}{2} 
+ 3 {f'}^2 + \frac{8ff'}{r} 
+ \frac{14f (1-f)}{r^2}\right]\psi'' \nonumber \\
&&
\hskip 1.5cm - \frac{2}{9}\left[\frac{ff'''}{2} 
+ 2f'f'' + \frac{9ff'' + 8{f'}^2}{r}
+ \frac{(5 - 17f)f'}{r^2}
+ \frac{6f(1-f)}{r^3}\right]\psi' \nonumber \\
&&
\hskip-.2cm +\, \frac{4f}{3} \left[\frac{f''}{2} 
- \frac{f'}{r} 
- \frac{1-f}{r^2}\right] \varphi''
+ \frac{2}{3}\left[ 2ff''' + \frac{f'f''}{2}
+ \frac{ff'' - {f'}^2}{r}
- \frac{f'(1+f)}{r^2}
- \frac{2f(1-f)}{r^3}\right]\varphi'\nonumber \\
&&
\hskip .5cm + \frac{2}{3} \left[ ff'''' 
+ \frac{f'f'''}{2}
- \frac{{f''}^2}{4}
+ \frac{f'f'' + 3 ff'''}{r}
- \frac{ff'' + {f'}^2}{r^2}
+ \frac{2ff'}{r^3}
+ \frac{1-f^2}{r^4}\right]\varphi\nonumber \\
&&
\hskip 1.5cm +\, \frac{4}{9}\left[\frac{{f''}^2}{4}
- \frac{f'f''}{r}
+ \frac{{f'}^2-f'' + ff''}{r^2}
+ \frac{2f'(1-f)}{r^3}
+ \frac{(1-f)^2}{r^4}\right]\,,
\end{eqnarray}
where primes denote differentaition with respect to $r$.

The radial pressures of the two stress tensors are:
\begin{eqnarray}
&&E^{\ r}_r= - f^2\varphi'\varphi'''
+ \frac{f^2}{2}{\varphi''}^2 
- \frac{4f}{3} \left(f' 
+ \frac{f}{r}\right) \varphi'\varphi''
+ \frac{1}{3} \left[ - 2 ff''
+ {f'}^2
- \frac{8ff'}{r}
+ \frac{2f(1+2f)}{r^2}\right]{\varphi'}^2\nonumber\\
&&
\hskip 1cm - \frac{2\kappa^2 p^2{f'}^2}{3f^2} 
+ \frac{8\kappa^2 p^2}{3rf} \left( f'
+ \frac{1-f}{r}\right)
+ \frac{f}{3} \left(f'
+ \frac{4f}{r}\right)\varphi'''\nonumber\\
&& 
+ \frac{1}{3} \left(- ff''
+ 2{f'}^2 
+ \frac{8ff'}{r} 
+ \frac{8f^2}{r^2}\right) \varphi''
+ \left[ \frac{ff'''}{3} 
+ \frac{2ff''}{r} 
+ \frac{2(3f-1)}{r^2} \left(f' 
- \frac{2}{3}\frac{f}{r}\right) \right] \varphi'\,,
\end{eqnarray}
and
\begin{eqnarray}
&&F^{\ r}_r = 
- f^2 (\varphi'\psi'''
+ \psi'\varphi''' 
- \varphi''\psi'')
- \frac{4f}{3}\left(f' 
+ \frac{f}{r}\right) \left(\varphi'\psi''
+ \psi'\varphi''\right)
- \frac{4\kappa^2pp'{f'}^2}{3f^2} \nonumber\\ 
&&
+ \frac{16\kappa^2pp'}{3rf} \left(f' 
+ \frac{1-f}{r}\right)
+ \frac{2}{3} \left[ {f'}^2 
- 2ff''
- \frac{8ff'}{r}
+ \frac{2f (1 + 2f)}{r^2} \right]\varphi'\psi'
+ \frac{f}{3} \left(f'
+ \frac{4f}{r}\right) \psi'''\nonumber\\
&&
\hskip .2cm + \frac{1}{3} \left(-ff'' 
+ 2{f'}^2
+ \frac{8ff'}{r} 
+ \frac{8{f}^2}{r^2}\right) \psi''
+ \left[\frac{ff'''}{3} 
+ \frac{2ff''}{r}
+ \frac{2(3f-1)f'}{r^2}
+ \frac{4f(1-3f)}{3r^3} \right]\psi'\nonumber\\
&&
\hskip 1cm + \frac{1}{3} \left[ f'f''
- \frac{2(ff'' + {f'}^2)}{r}
+ \frac{2(3f-1)f'}{r^2}
+ \frac{4f(1-f)}{r^3}\right] \varphi'\nonumber\\
&&
\hskip .5cm + \frac{1}{3} \left[f'f''' 
- \frac{{f''}^2}{2}
+ \frac{2(f'f''- ff''')}{r}
- \frac{2(ff'' + {f'}^2)}{r^2}
+ \frac{4ff'}{r^3} 
+ \frac{2(1-f^2)}{r^4}\right]\varphi\,.
\end{eqnarray}
Finally the tangential pressures of the two tensors are:
\begin{eqnarray}
&&E^{\ \theta}_{\theta} = 
\frac{f^2}{3}\left( \varphi'\varphi'''
- \frac{{\varphi''}^2}{2} \right)
+ \frac{4f^2}{3r}\, \varphi'\varphi''
+ \frac{f'}{3} \left(-f' 
+ \frac{4f}{r}\right) {\varphi'}^2
+ \frac{2\kappa^2p^2}{3f^2} \left(2ff'' 
- {f'}^2\right) 
\nonumber \\ &&
- \frac{2f^2}{3r}\, \varphi'''
+ \frac{2f}{9} \left(f''
- \frac{2f'}{r}
+ \frac{7 - 13f}{r^2} \right)\varphi''
+ \frac{2}{9} \left[ -\frac{ff'''}{2}
+ f'f''
+ \frac{4{f'}^2}{r}
+ \frac{(7 - 22f)f'}{r^2}
+ \frac{6f^2}{r^3} \right] \varphi'
\nonumber \\ &&
+ \frac{4}{9} \left[ \frac{ff''''+ f'f'''}{2}
+ \frac{3ff'''+ 2f'f''}{r}
+ \frac{4ff'' - 3f'' + 2{f'}^2} {r^2}
+ \frac{2f'(1-2f)}{r^3} 
+ \frac{2f(f-1)}{r^4} \right]\,,
\end{eqnarray}
and
\begin{eqnarray}
&&F^{\ \theta}_{\theta} = \frac{f^2}{3}\left(\varphi'\psi'''
+ \psi'\varphi''' - \varphi''\psi''\right)
+ \frac{4f^2}{3r} \left(\psi'\varphi''
+ \varphi'\psi''\right)
+ \frac{2f'}{3} \left( -f' + \frac{4f}{r}\right) \varphi'\psi'
\nonumber \\ &&
\hskip .5cm + \frac{4\kappa^2pp'}{3f^2}\left(2ff'' - {f'}^2\right) 
- \frac{2f^2}{3r}\psi'''
+ \frac{2f}{9} \left(f'' 
- \frac{2f'}{r}
+ \frac{7 - 13f}{r^2}\right)\psi''
\nonumber \\ &&
+ \frac{2}{9}\left[- \frac{ff'''}{2} + f'f''
+ \frac{4{f'}^2}{r}
+ \frac{(7-22f)f'}{r^2}
+ \frac{6f^2}{r^3}\right] \psi'
+ \frac{f}{3} \left[-f'' + \frac{2f'}{r}
+ \frac{2(1-f)}{r^2}\right] \varphi''
\nonumber \\ &&
\hskip 1cm + \frac{2}{3} \left[ -ff''' - \frac{f'f''}{2}
+ \frac{{f'}^2}{r} 
+ \frac{(1-f)f'}{r^2}\right] \varphi'
\nonumber \\ &&
+ \frac{2}{3} \left[-\frac{(ff'''' + f'f''')}{2}
+ \frac{{f''}^2}{4}
- \frac{(ff''' + f'f'')}{r}
+ \frac{ff'' + {f'}^2}{r^2} 
- \frac{2ff'}{r^3} 
+ \frac{f^2 - 1}{r^4}\right] \varphi
\nonumber \\ &&
+ \frac{4}{9} \left[ \frac{{f''}^2}{4}
- \frac{f'f''}{r}
+ \frac{(f-1) f'' + {f'}^2}{r^2}  
+ \frac{2(1-f)f'}{r^3} 
+ \frac{(f-1)^2}{r^4} \right]\,.
\end{eqnarray}
The flux components $E_t^{\ r}$ and $F_t^{\ r}$ are determined by the
conservation Eqn.,
\begin{equation}
\frac{\partial}{\partial r}\, T_t^{\ r} + \frac{2}{r} \, T_t^{\ r} = 0
\end{equation}
to be proportional to $1/r^2$ with a proportionality constant given by
the luminosity $L$ as in (\ref{lumflux}) of the text. For the general
surface gravity $\kappa$, $T_t^{\ r} = - L/4\pi r^2$ with
\begin{equation}
L = 16\pi \kappa^2 \left( bpq' + bp'q + b'pq\right)\,.
\end{equation}

\begin{subfigures}
\begin{figure}[p]
\vskip -0.1in
\hskip -0.4in
\includegraphics[angle=0,height=3.6in,width=5.4in]
{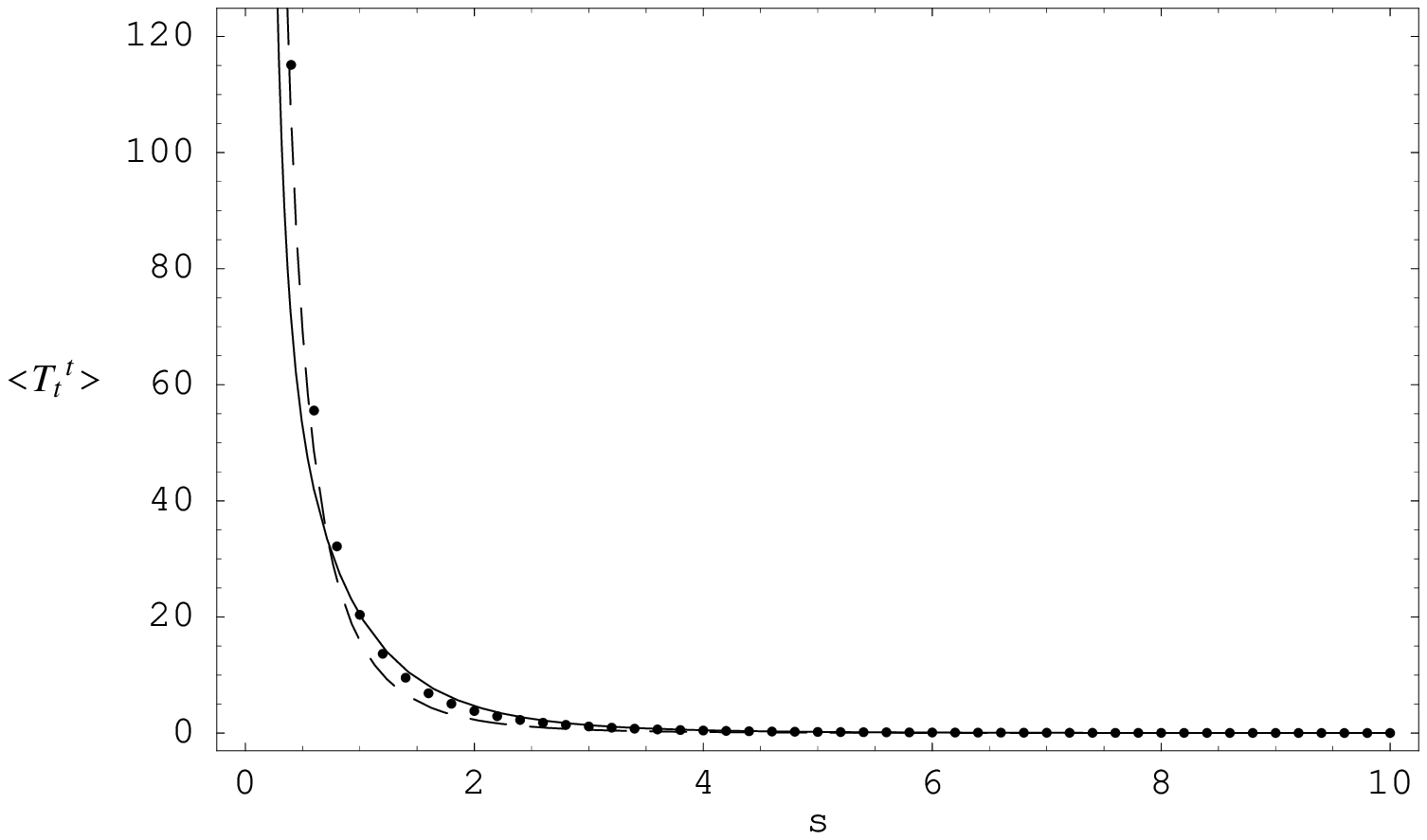}
\vskip -.2in
\caption{The expectation value $\langle T_t^{\ t} \rangle$ of a conformal 
scalar field in the Boulware state in Schwarzschild spacetime, as a function 
of $s=\frac{r-2M}{M}$ in units of $\pi^2 T_H^4/90$. The solid curve is Eq. 
(\ref{Tanom}) with (\ref{Boulcond}) and $c_{_H} = -\frac{7}{20}, d_{_H} 
= \frac{55}{84}$, the dashed curve is the analytic approximation of 
\cite{BroOtt}, and the points are the numerical results of \cite{JenMcOtt}.}
\label{fig:TBOtt}
\vskip 0.7in
\hskip -0.4in
\includegraphics[angle=0,height=3.6in,width=5.4in]
{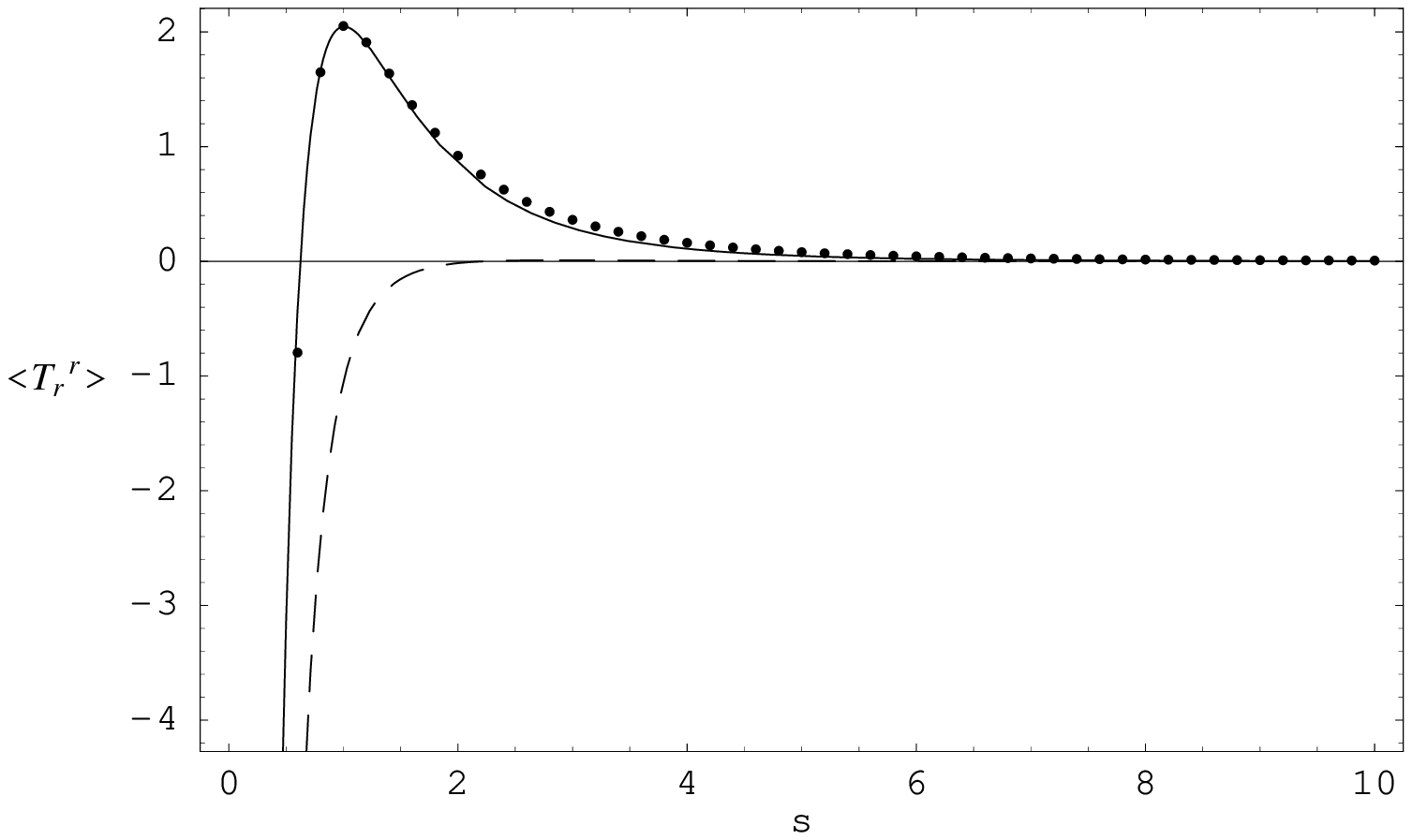}
\vskip -.2in
\caption{The radial pressure $\langle T_r^{\ r} \rangle$ of a 
conformal scalar field in the Boulware state in Schwarzschild spacetime.
The axes and solid and dashed curves and points are as in Fig. \ref{fig:TBOtt}.}
\label{fig:TBOrr}
\vskip -0.5in
\end{figure}

\begin{figure}[t]
\vskip -.1in
\hskip -0.4in
\includegraphics[angle=0,height=3.6in,width=5.4in]
{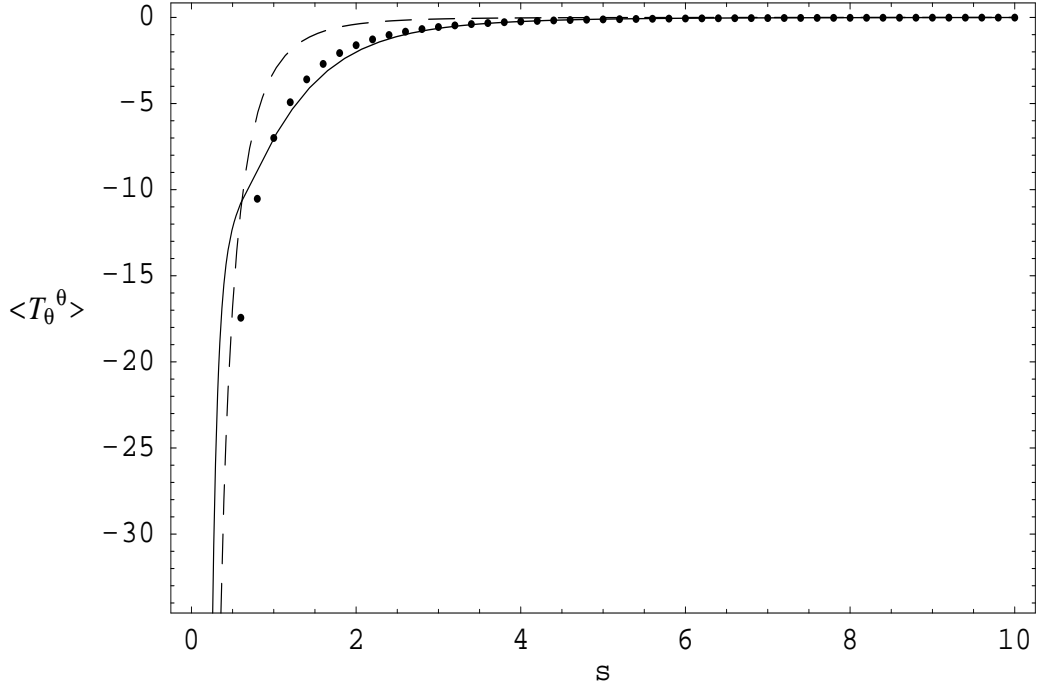}
\vskip -.2in
\caption{The tangential pressure $\langle T_\theta^{\ \theta} \rangle$ of 
a conformal scalar field in the Boulware state in Schwarzschild spacetime.
The axes and solid and dashed curves and points are as in Fig. \ref{fig:TBOtt}.}
\label{fig:TBOthth}
\end{figure}
\end{subfigures}

\begin{subfigures}
\begin{figure}[b]
\vskip .4in
\hskip -0.4in
\includegraphics[angle=0,height=3.6in,width=5.4in]
{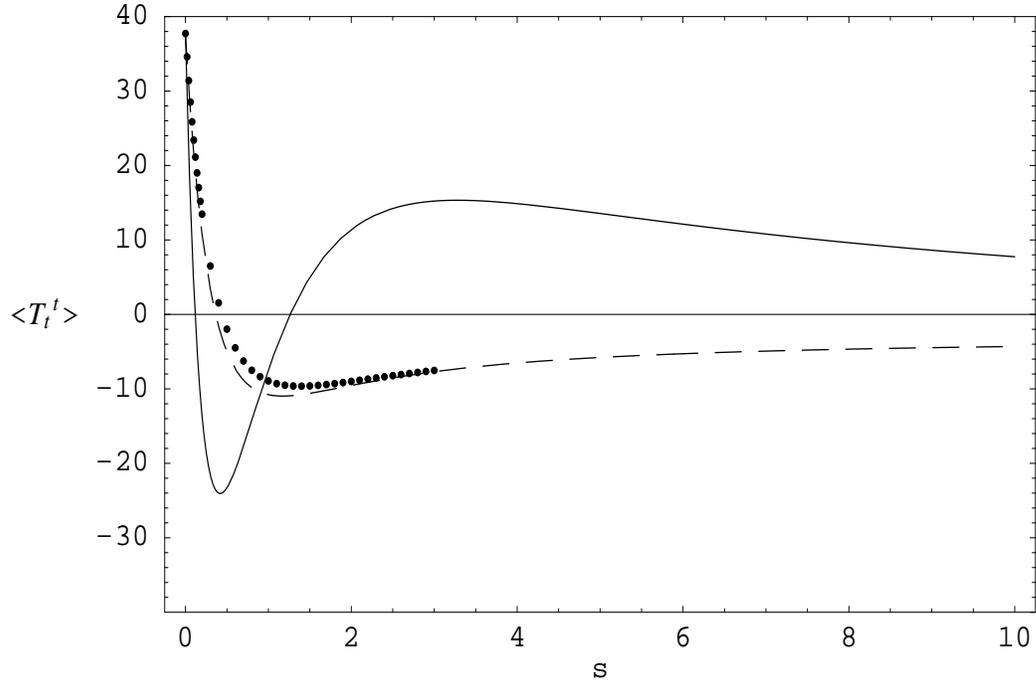}
\vskip -.3in
\caption{The expectation value $\langle T_t^{\ t} \rangle$ of a conformal 
scalar field in the Hartle-Hawking state in Schwarzschild spacetime as a 
function of $s=\frac{r-2M}{M}$ in units of $\pi^2 T_H^4/90$. The solid curve 
is Eq. (\ref{Tanom}) with $c_{\infty}=0.035, c_{_H}=-1.5144, p=1.5144, q=q'=2,
d_\infty=1.0262, d_{_H}=1.0096, p'=-1.0096$, the dashed curve is the analytic 
approximation of \cite{FroZel} and the data points are the numerical results 
of \cite{How}.}
\label{fig:THH0tt}
\vskip -.3in
\end{figure}

\begin{figure}[p]
\vskip -.1in
\hskip -0.4in
\includegraphics[angle=0,height=3.6in,width=5.4in]
{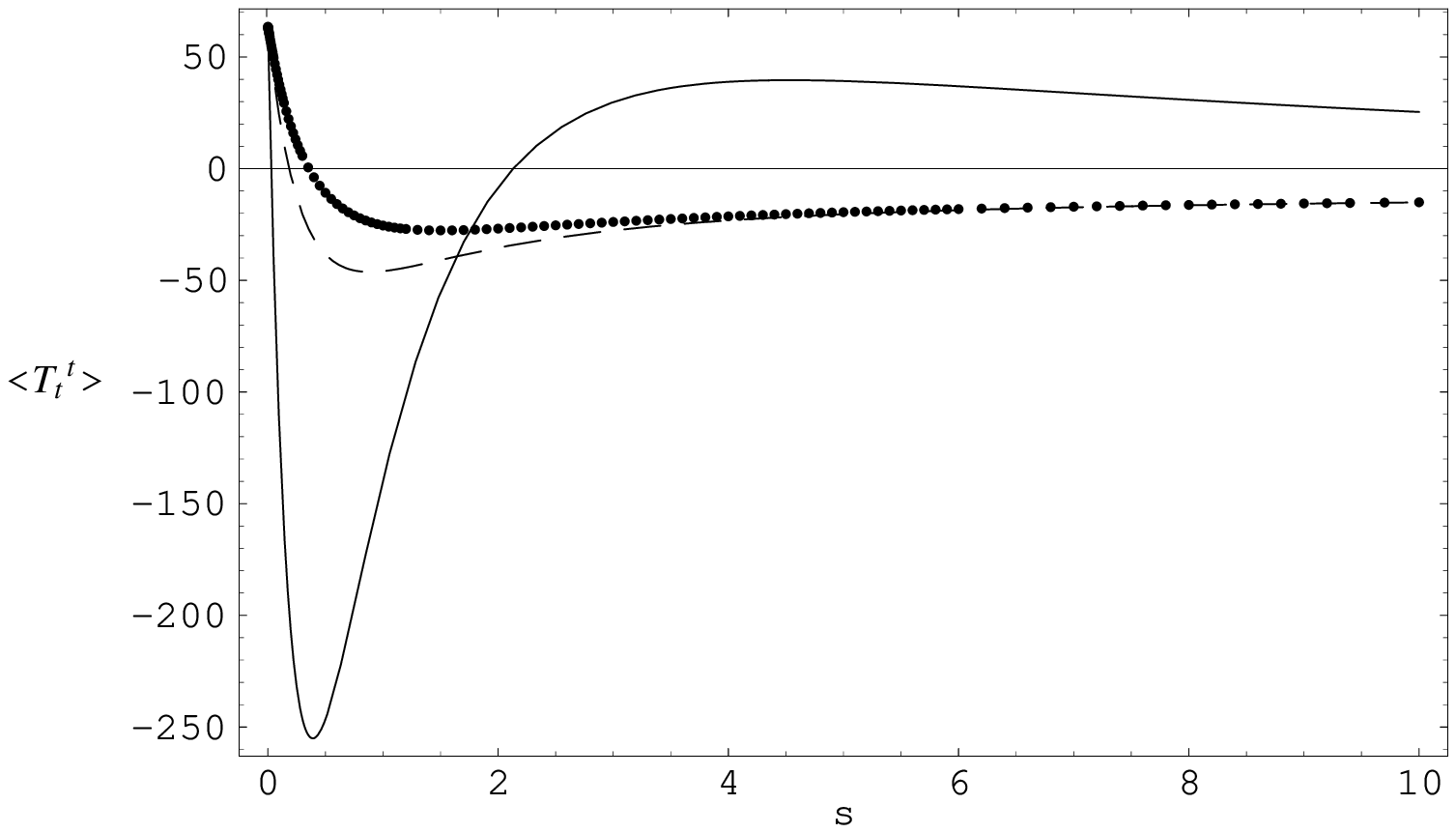}
\vskip -.2in
\caption{The expectation value $\langle T_t^{\ t} \rangle$ of a massless 
Dirac field in the Hartle-Hawking state in Schwarzschild spacetime. The solid 
curve is Eq. (\ref{Tanom}) with $c_{\infty}=0.035, c_{_H}=-1.571 = -p, q=q'=2,
d_\infty=0.6059, d_{_H}=0.6109 = -p'$, the dashed curve is the analytic 
approximation of \cite{FroZel} and the data points are the numerical results of 
\cite{CHOAG}.}
\label{fig:THH12tt}
\vskip .5in
\hskip -0.4in
\includegraphics[angle=0,height=3.6in,width=5.4in]
{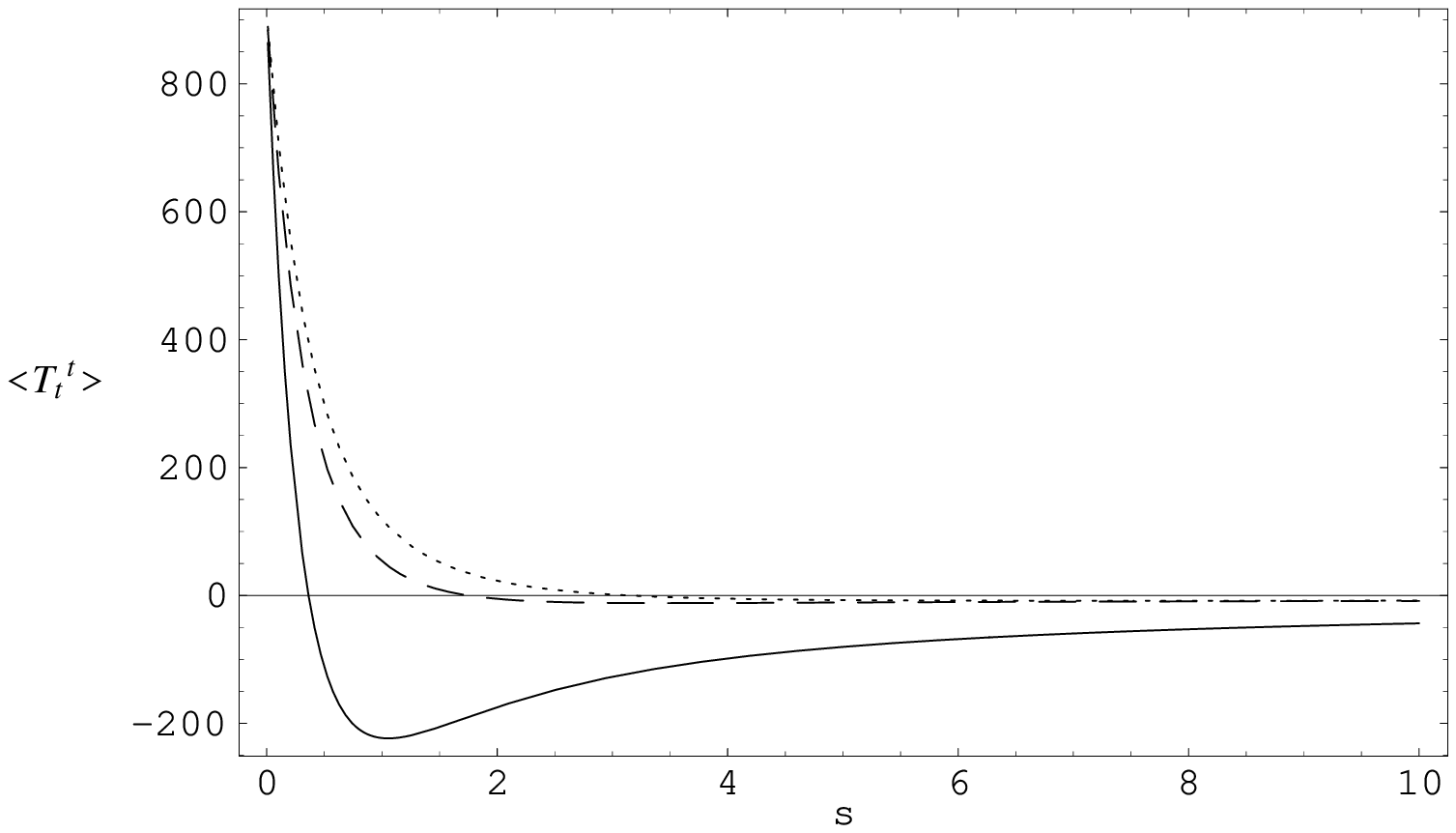}
\vskip -.2in
\caption{The expectation value $\langle T_t^{\ t} \rangle$ of the 
electromagnetic field in the Hartle-Hawking state in Schwarzschild spacetime. 
The solid curve is eq. (\ref{Tanom}) with $c_{\infty}=-0.0237, c_{_H}=1.0584 = p, 
q=q'=2, d_\infty = -0.2716, d_{_H} = 0.7644 = p'$, and the dashed curve is the 
accurate analytic fit of \cite{JenOtt} to the numerical results of the same authors.}
\label{fig:THH1tt}
\vskip -.6in
\end{figure}
\end{subfigures}

\begin{subfigures}
\begin{figure}[p]
\vskip -.1in
\hskip -.4in
\includegraphics[angle=0,height=3.6in,width=5.4in]
{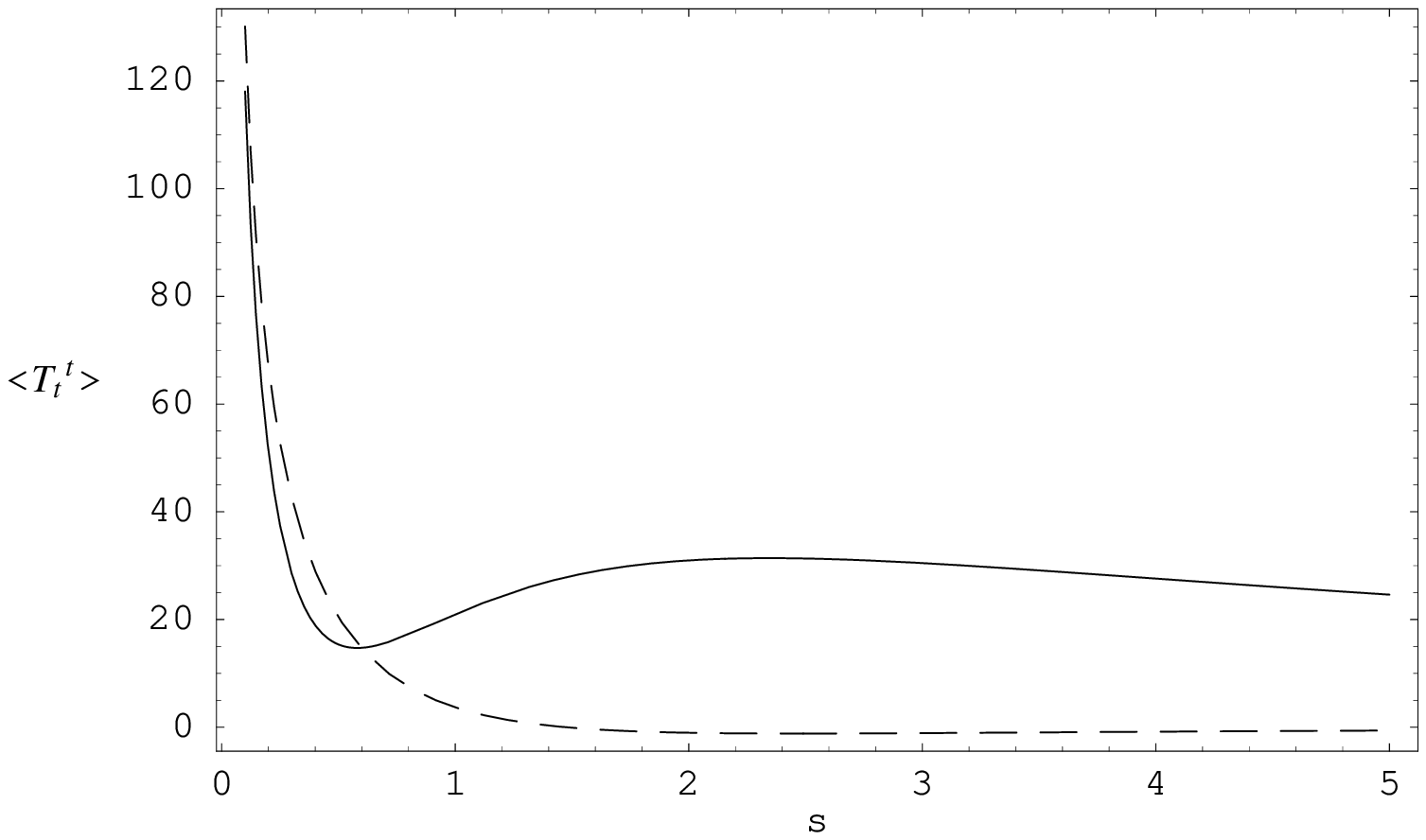}
\vskip -.3in
\caption{The expectation value $\langle T_t^{\ t} \rangle$ of a massless, 
conformal field in the Unruh state in Schwarzschild spacetime. The solid curve 
is Eq. (\ref{Tanom}) with $c_{\infty}= 0, c_{_H} = -1.6500, p = 1.9192,
q=q'=2, d_\infty = 1.4441, d_{_H} = 1.3244, p' = -1.0552$, and the dashed curve 
is the polynomial approximation of \cite{Vis} which is an accurate fit to the 
numerical results of \cite{Elst}.}
\label{fig:TU0tt}
\vskip .5in
\hskip -.4in
\includegraphics[angle=0,height=3.6in,width=5.4in]
{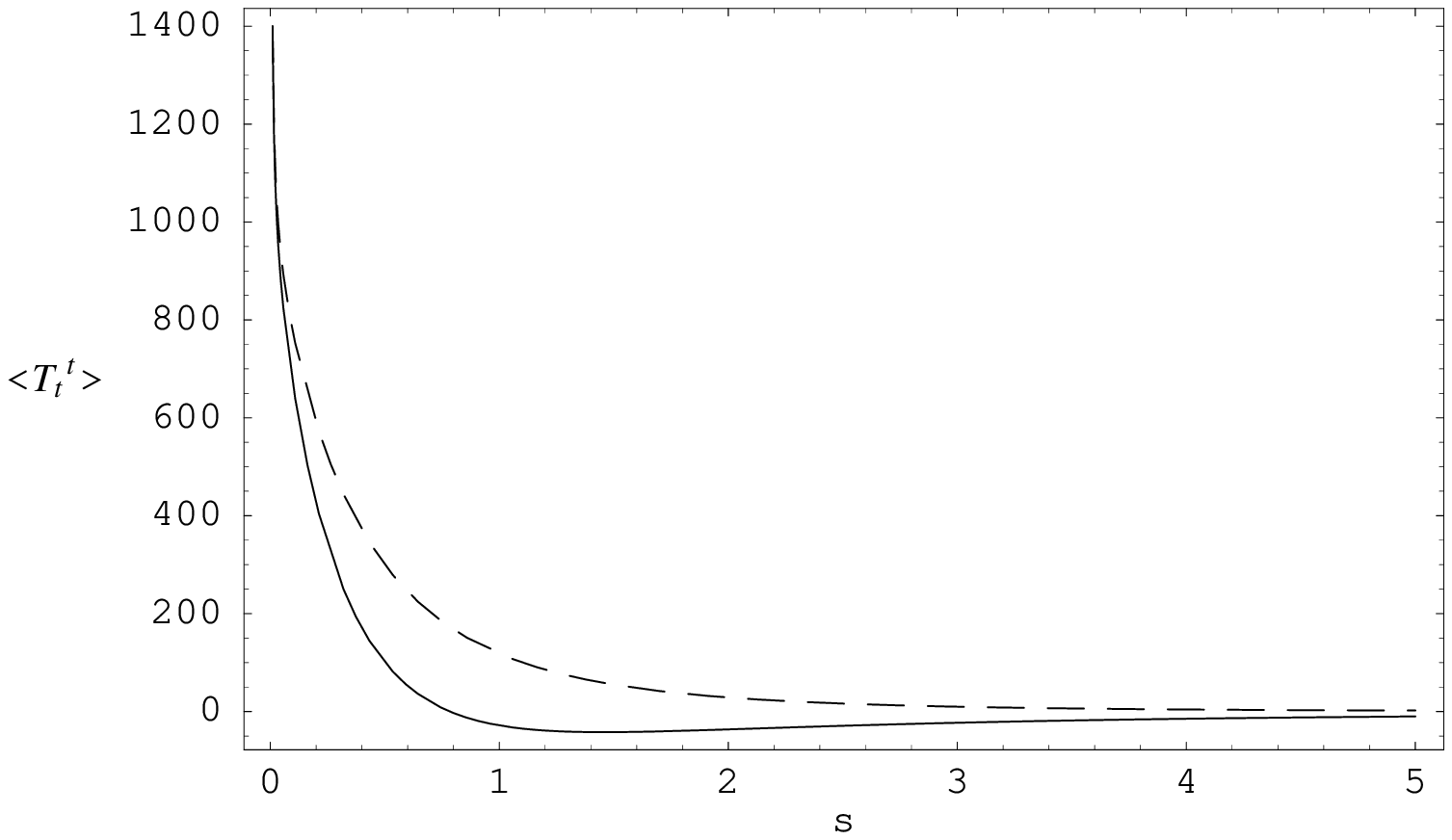}
\vskip -.2in
\caption{The expectation value $\langle T_t^{\ t} \rangle$ of the 
electromagnetic field in the Unruh state in Schwarzschild spacetime. The solid 
curve is eq. (\ref{Tanom}) with $c_{\infty} = 0, c_{_H} = 0.5594, p = -0.4984,
q=q'=2, d_\infty = -0.2595, d_{_H} = 0.4125, p' = -.3514$, and the dashed curve 
is the polynomial approximation of \cite{Matyj} which is an accurate fit to the 
numerical results of \cite{JenMcOttBR}.}
\label{fig:TU1tt}
\vskip -.6in
\end{figure}
\end{subfigures}

\end{document}